\documentclass[12pt]{article}
\usepackage{graphicx,psfrag,color}

\usepackage{amsmath,amssymb}
\numberwithin{equation}{section}
\allowdisplaybreaks

\usepackage{subfig}
\usepackage{braket}
\usepackage{cite}

\newcommand{\p}{\partial}
\newcommand{\ra}{\rightarrow}

\begin{document}

\begin{titlepage}

\begin{flushright}
TUM-HEP-920/13\\
TTK-13-27\\
SFB/CPP-13-111 \\
December 16, 2013
\end{flushright}

\vskip1.5cm
\begin{center}
\Large\bf\boldmath P-wave contribution to third-order 
top-quark pair production near threshold 
\end{center}

\vspace{1cm}
\begin{center}
{\sc M.~Beneke}$^a$, {\sc J.~Piclum}$^{a,b}$, and {\sc T.~Rauh}$^a$\\[5mm]
  {\it $^a$ Physik Department T31, James-Franck-Stra\ss e~1,}\\
  {\it Technische Universit\"at M\"unchen, D--85748 Garching, Germany}\\[0.1cm]
  {\it $^b$ Institut f{\"u}r Theoretische Teilchenphysik und Kosmologie,}\\
  {\it RWTH Aachen University, D--52056 Aachen, Germany}\\[0.3cm]
\end{center}

\vspace{2cm}
\begin{abstract}
\noindent 
The next-to-leading order (NLO) P-wave Coulomb Green function contributes at 
third-order to top-pair production in $e^+e^-$ collisions near 
threshold. In this paper we compute the NLO P-wave Green function in 
dimensional regularization, as required for a consistent combination 
with non-resonant production of the $W^+ W^-b\bar b$ final state, and present a  
phenomenological analysis of the P-wave contribution. We further briefly 
discuss squark production near threshold and top-pair production in 
$\gamma\gamma$ collisions, where no S-wave contribution is present, 
and the P-wave thus constitutes the dominant production process.
\end{abstract}
\end{titlepage}

\pagenumbering{roman}
\newpage
\pagenumbering{arabic}

\section{Introduction\label{sec:introduction}}

A future high-energy electron-positron collider will allow a very 
precise measurement of the
top-antitop production cross section near threshold. From the
threshold scan several standard model parameters, like top-quark mass,
width, and Yukawa coupling, can be extracted with high precision. It
was found in several
studies~\cite{Martinez:2002st,Seidel:2013sqa,Horiguchi:2013wra} that
the top mass can be determined with an uncertainty well below
$100$~MeV. Contrary to direct reconstructions at hadron colliders,
there is no ambiguity in relating the result to a precisely defined 
mass parameter. To achieve this level of accuracy requires 
precise theoretical predictions for the threshold cross section. The
challenge is that conventional perturbation expansions in the strong 
coupling $\alpha_s$ fail for threshold production, since it involves 
multiple scales. In terms of the mass $m_t$ and velocity $v$ of the 
top-quark these are the hard scale $m_t$, the soft scale $m_tv$, and 
the ultrasoft scale $m_tv^2$. At the ultrasoft scale, the colour-Coulomb 
force is non-perturbatively strong. In terms of Feynman diagrams, 
this implies that when the velocity is of order of the 
strong coupling, Coulomb singularities of the form $(\alpha_s/v)^n$ 
have to be summed to all orders. This can be achieved by successively 
integrating out the hard and soft scale leading to the non-relativistic 
QCD (NRQCD)~\cite{Thacker:1990bm,Lepage:1992tx,Bodwin:1994jh} and potential
NRQCD
(PNRQCD)~\cite{Pineda:1997bj,Pineda:1997ie,Beneke:1998jj,Beneke:1999qg,Brambilla:1999xf} effective field theories, respectively. Within this framework, 
described in detail for top-quark pair production near threshold 
in \cite{Beneke:2013PartI}, the dominant contribution from
the S-wave correlation function has been computed at
next-to-next-to-next-to-leading order
(NNNLO)~\cite{Beneke:2005hg,Beneke:2008ec,Beneke:2008cr,Beneke:2013PartII}.

The axial-vector coupling of the top quark to the $Z$ boson gives 
rise to a P-wave contribution to the top-pair production cross section. 
In this work we compute the corresponding P-wave PNRQCD correlation 
function. Being suppressed by $v^2$ relative to the S-wave it contributes
only starting from NNLO. The complete NNNLO calculation of the threshold 
correction therefore requires a NLO calculation of the P-wave 
correlation function, which we perform here in dimensional regularization. 
Some results for the P-wave Green function were
already obtained 
in~\cite{Bigi:1991mi,Penin:1998ik,Penin:1998mx,Kuhn:1999hw}, but
none of these computations were performed in dimensional
regularization. Dimensional regularization is, however, required 
for the following reason: The imaginary part of the P-wave Green function, 
which is relevant for the cross section, is
divergent already at leading order in the non-relativistic expansion,
if the finite width of the top quark is included. This divergence and
the resulting scheme dependence cancel only when non-resonant 
corrections to the process $e^+ e^-\to W^+ W^-b\bar b$ are added. 
The separation of resonant and non-resonant contributions can be 
performed consistently in unstable-particle effective field 
theory~\cite{Beneke:2003xh, Beneke:2004km}, in which the non-resonant 
terms appear as a hard region. The corresponding diagrams are 
computed as usual in dimensional regularization as has 
already been done in~\cite{Beneke:2010mp,Jantzen:2013gpa}. Consistency then 
requires that the non-relativistic, resonant part is also computed 
in dimensional regularization. We emphasize that to determine the 
top-quark mass precisely it is necessary to compute  
the process $e^+ e^-\to W^+ W^-b\bar b$ including non-resonant terms, 
since the top mass is ultimately determined from the rise of the 
cross section near threshold, where non-resonant effects are 
important~\cite{Beneke:2010mp}.

The outline of the paper is as follows.
In Sec.~\ref{sec:theory} we give a brief overview of the framework
of the calculation. A detailed discussion can be found
in~\cite{Beneke:2013PartI,Beneke:2013PartII}, however, we repeat some
of the formulas given there in order to make the present paper
self-contained. The P-wave Green function up to NLO is computed in
Sec.~\ref{sec:computation} with some technical details relegated to 
the appendix. In Sec.~\ref{sec:crosssection} a numerical 
analysis of the P-wave contribution to top-pair production at
threshold is presented. The absolute size of the P-wave contribution 
is rather small due to the small axial couplings of the top quark.
We therefore also  discuss the P-wave dominated (s)top threshold
production processes $\gamma\gamma\ra t\bar{t}$ with different photon
helicities and $e^+e^-\ra\tilde{t}\tilde{\bar{t}}$ in
Sec.~\ref{sec:photon} and Sec.~\ref{sec:susy}, respectively, where 
P-wave production is the dominant production mechanism. We
conclude in Sec.~\ref{sec:conclusion}.

\section{Effective theory setup\label{sec:theory}}

The production of a top pair in $e^+e^-$ annihilation is mediated by
photons and $Z$ bosons. While the coupling of photons to fermions is
purely vector-like the $Z$ boson couples to vector currents and 
axial-vector currents with the respective strengths
\begin{equation}
  v_f=\frac{T_3^f-2e_f\sin^2\theta_w}{2\sin\theta_w\cos\theta_w},
  \hspace{1cm}
  a_f=\frac{T_3^f}{2\sin\theta_w\cos\theta_w},
  \label{eq:Zffcoupling}
\end{equation}
where $\theta_w$ is the Weinberg angle, $e_f$ the electric charge and
$T_3^f$ the third component of the weak isospin of the fermion
$f$. For the vector current $j_\mu^{(v)}=\bar{t}\gamma_\mu t$ and the
axial-vector current $j_\mu^{(a)}=\bar{t}\gamma_\mu\gamma_5 t$ we
define the two-point functions
\begin{equation}
  \begin{aligned}
    \Pi_{\mu\nu}^{(X)}(q^2) =& \hspace*{0.15cm}
    i\int d^dx\,e^{iq\cdot x}\Braket{0|T\left[j_\mu^{(X)}(x)j_\nu^{(X)}(0)\right]|0}\\
    =& \hspace*{0.15cm}
(q_\mu q_\nu-q^2g_{\mu\nu})\,\Pi^{(X)}(q^2)+q_\mu q_\nu\Pi_L^{(X)}(q^2).
  \end{aligned}
  \label{eq:twopointfct}
\end{equation}
We denote by $R=\sigma_{X\bar{t}t}/\sigma_0$ the inclusive $\bar{t}t$
production cross section $\sigma_{X\bar{t}t}=\sigma(e^+e^-\ra
t\bar{t}X)$ normalized to the high-energy limit of the $\mu^+\mu^-$
production cross section
$\sigma_0=4\pi\alpha_{\text{em}}^2/(3q^2)$. It can be related to the
imaginary part of the two-point functions by the optical theorem
\begin{equation}
  \begin{aligned}
    R = 12\pi\text{ Im} \bigg[&
    e_t^2\Pi^{(v)}(q^2)-\frac{2q^2}{q^2-M_Z^2}v_ev_te_t\Pi^{(v)}(q^2)\\
    &+\left(\frac{q^2}{q^2-M_Z^2}\right)^2(v_e^2+a_e^2)(v_t^2\Pi^{(v)}(q^2)+a_t^2\Pi^{(a)}(q^2))
    \bigg].
  \end{aligned}
  \label{eq:Rratio}
\end{equation}

Near the production threshold $s\equiv q^2\approx4m_t^2$ the usual
perturbation theory in $\alpha_s$ breaks down, because the velocity
$v$ of the top quark is of the same order as the strong coupling and
contributions that scale as $(\alpha_s/v)^k$ have to be summed to
all orders. Instead, in non-relativistic perturbation theory, one expands 
in $\alpha_s$ and $v$ around the non-perturbative solution 
that sums these terms. Explicitly, the re-organized expansion takes the form
\begin{eqnarray}
  \begin{aligned}
    R \sim v \sum\limits_k \left(\frac{\alpha_s}{v}\right)^k
    \{& 1(\text{LO}); \alpha_s,v(\text{NLO});
    \alpha_s^2,\alpha_sv,v^2(\text{NNLO}); \\
    &\alpha_s^3,\alpha_s^2v,\alpha_sv^2,v^3(\text{NNNLO});\dots\}.
  \end{aligned}
  \label{eq:Rexpansion}
\end{eqnarray}

The computation of the vector current spectral function
$\Pi^{(v)}(q^2)$ to NNNLO is summarized in~\cite{Beneke:2013PartI,Beneke:2013PartII}. This
work focuses on the axial-vector current spectral function
$\Pi^{(a)}(q^2)$. In a first step the hard modes are integrated
out. Matching of the axial-vector current to NRQCD yields an
expansion
\begin{equation}
 j^{(a)k}=\frac{c_a}{2m_t}\psi^\dagger\left[\sigma^k,(-i)\boldsymbol{\sigma}\cdot\mathbf{D}\right]\chi+\dots,
 \label{eq:axialvectorcurrent}
\end{equation}
where bold face letters and latin indices refer to $d-1$ dimensional vectors, 
and $d=4-2\epsilon$ is the space-time dimension. 
The hard matching coefficient $c_a$ is given by~\cite{Kniehl:2006qw}\footnote{
The $\mathcal{O}(\epsilon)$ term will be needed later on and was not given 
in~\cite{Kniehl:2006qw}. For $\gamma_5$ the naive anti-commuting (NDR) scheme 
is employed.}
\begin{equation}
 c_a=1-4 C_F\,\frac{\alpha_s}{4\pi} \left[1-\epsilon\ln\frac{m_t^2}{\mu^2} 
+\mathcal{O}(\epsilon^2)\right] +\mathcal{O}(\alpha_s^2).
  \label{eq:ca}
\end{equation}
We observe that~\eqref{eq:axialvectorcurrent} contains a covariant
derivative. Thus the axial-vector current spectral function is
suppressed by the well-known P-wave factor of $v^2$ compared to the S-wave 
and contributes to $R$ only starting from NNLO. The two-point function
takes the form
\begin{equation}
\label{eq:PiaNRQCD}
  \begin{aligned}
   \Pi^{(a)}(q^2) =&  \hspace*{0.15cm}\frac{1}{(d-1)q^2}\,\Pi_{ii}^{(a)}(q^2)\\
    =&  \hspace*{0.15cm}\frac{N_c c_a^2}{8m_t^4}\frac{i}{2N_c(d-1)}
    \int d^dx\,e^{iEx^0}
    \Braket{0|T\left([\psi^\dagger\Gamma^i\chi]^\dagger(x)
        [\psi^\dagger\Gamma^i\chi](0)\right)|0}_{\text{NRQCD}}+\dots,
 \end{aligned}
\end{equation}
where
$\Gamma^k=(-i)\left[\sigma^k,\boldsymbol{\sigma}\cdot\mathbf{D}\right]$. A
further matching to potential NRQCD integrates out the soft modes and
the potential gluons and light quarks. To the required order the
PNRQCD Lagrangian reads
\begin{equation}
 \begin{aligned}
  \mathcal{L}_{\text{PNRQCD}}=&\psi^\dagger\left(i\p_0+g_sA_0(t,\mathbf{0})+\frac{\boldsymbol{\p}^2}{2m}\right)\psi+\chi^\dagger\left(i\p_0+g_sA_0(t,\mathbf{0})-\frac{\boldsymbol{\p}^2}{2m}\right)\chi\\
  &+\int d^{d-1}\mathbf{r}\left[\psi_a^\dagger\psi_b\right](x+\mathbf{r})V_{ab;cd}(\mathbf{r},\p)\left[\chi_c^\dagger\chi_d\right](x),
 \end{aligned}
\label{eq:LPNRQCD}
\end{equation}
where $\psi$ ($\chi$) denotes the potential quark (antiquark) field. The
coupling to the ultrasoft gluon field 
$A_0(t,\mathbf0)$ can be removed by a field redefinition,
which, in general, modifies the external current~\cite{Beneke:2010da}, 
but cancels for the colour-singlet currents relevant to production 
through photons and $Z$ bosons. 
Henceforth, ultrasoft gluons can be ignored, since they contribute only at higher
order. 
Since the top pair is produced in a colour-singlet state, only
the colour-singlet projection
\begin{equation}
 V(\mathbf{p},\mathbf{p}')=\frac{1}{N_c}\delta_{bc}\delta_{da}V_{ab;cd}(\mathbf{p},\mathbf{p}')
\end{equation}
of the general potential is required. To the considered order it
consists purely of the Coulomb potential
\begin{equation}
 V(\mathbf{p},\mathbf{p}')=-\frac{4\pi C_F\alpha_s}{\mathbf{q}^2}\left[\mathcal{V}_C^{(0)}+\frac{\alpha_s}{4\pi}\mathcal{V}_C^{(1)}+\mathcal{O}(\alpha_s^2)\right]+\dots,
 \label{eq:VCoulomb}
\end{equation}
where the $d$ dimensional coefficients of the LO and NLO Coulomb potential are
\begin{align}
 \mathcal{V}_C^{(0)}=&\hspace*{0.15cm}1,\\
 \mathcal{V}_C^{(1)}=&\hspace*{0.15cm}\left[\left(\frac{\mu^2}{\mathbf{q}^2}\right)^\epsilon-1\right]\frac{\beta_0}{\epsilon}+\left(\frac{\mu^2}{\mathbf{q}^2}\right)^\epsilon a_1(\epsilon),
 \label{eq:VCoulombcoefficients}
\end{align}
with
\begin{equation}
 a_1(\epsilon)=\Big(C_A[11-8\epsilon]-4T_Fn_f\Big)\frac{e^{\gamma_E\epsilon}\Gamma(1-\epsilon)\Gamma(2-\epsilon)\Gamma(\epsilon)}{(3-2\epsilon)\Gamma(2-2\epsilon)}-\frac{\beta_0}{\epsilon}.
\end{equation}
They are required up to order $\epsilon$ for reasons that will become
clear later. The LO Coulomb potential contributes to the same order as
the leading kinetic terms in the PNRQCD Lagrangian and thus has to be treated
non-perturbatively, while the NLO correction $ \mathcal{V}_C^{(1)}$ is a 
perturbation. Thus the LO Lagrangian describes the propagation of 
quark-antiquark pairs, where ladder diagrams with exchange of an 
arbitrary number of potential gluons between the quark-antiquark pair have been
resummed. The quark-antiquark pair propagator
$\tilde{G}_0(\mathbf{p},\mathbf{p}';E)$ satisfies the $d$-dimensional
Lippmann-Schwinger equation
\begin{equation}
 \begin{aligned}
  &\left(\frac{\mathbf{p}^2}{m_t}-E\right)
\tilde{G}_0(\mathbf{p},\mathbf{p}';E)-\tilde{\mu}^{2\epsilon}
\int\frac{d^{d-1}\mathbf{k}}{(2\pi)^{d-1}}
\frac{4\pi C_F\alpha_s}{\mathbf{k}^2}\,
\tilde{G}_0(\mathbf{p}-\mathbf{k},\mathbf{p}';E)\\
  & = (2\pi)^{d-1}\delta^{(d-1)}(\mathbf{p}-\mathbf{p}'),
 \end{aligned}
 \label{eq:LippmannSchwinger}
\end{equation}
where $E=\sqrt{s}-2m_t$ denotes the energy relative to the production
threshold, $m_t$ the top-quark pole mass, 
and the scale $\tilde\mu=\mu \,[e^{\gamma_E}/(4\pi)]^{1/2}$ has
been chosen such that the minimal subtraction of $1/\epsilon$ 
poles corresponds to
the $\overline{\text{MS}}$ rather than the MS scheme. Here and in the
following the tilde is used to indicate that the Green function is
given in momentum space. Its Fourier transform
\begin{equation}
 G_0(\mathbf{r},\mathbf{r}';E)=
\int\frac{d^{d-1}\mathbf{p}}{(2\pi)^{d-1}}\int\frac{d^{d-1}\mathbf{p}'}
{(2\pi)^{d-1}}\,e^{i\mathbf{p}\cdot\mathbf{r}}e^{-i\mathbf{p}'\cdot\mathbf{r}'}
\tilde{G}_0(\mathbf{p},\mathbf{p}';E)
\end{equation}
is the solution to the Schr\"odinger equation
\begin{equation}
 \left(-\frac{\nabla_{(r)}^2}{m_t}-\frac{C_F\alpha_s}{r}-E\right)
G_0(\mathbf{r},\mathbf{r}';E)=\delta^{(3)}(\mathbf{r}-\mathbf{r}')
 \label{eq:Schroedinger}
\end{equation}
in four dimensions. An expression for general $d$ is not available,
but for $d=4$ several representations are
known~\cite{Wichmann:1961,Schwinger:1964zzb,Voloshin:1985bd,Voloshin:1979uv}. We
find it convenient to use the integral representation 
from~\cite{Wichmann:1961}, 
\begin{equation}
  \begin{aligned}
    G_0(\mathbf{r},\mathbf{r}';E) =&\hspace*{0.15cm}
    \sum\limits_{l=0}^\infty\, 
    (2 l+1)\, P_l\!\left(\frac{\mathbf{r}\cdot\mathbf{r}'}{rr'}\right)
    \,\frac{m_t p}{2\pi} \frac{(2pr)^{l}(2 pr')^l}
    {\Gamma(l+1+\lambda)\Gamma(l+1-\lambda)}\,
    \\
    &\times\int\limits_0^1dt\int\limits_1^\infty ds \left[s(1-t)\right]^{l+\lambda}
    \left[t(s-1)\right]^{l-\lambda}\exp\left\{-p\,[r'(1-2t)+r(2s-1)]\right\},
  \end{aligned}
  \label{eq:LOGFintegralrep}
\end{equation}
with $r'<r$ and $p=\sqrt{-m_tE}$, and $P_l(z)$ the Legendre polynomials. 
We also introduced the variable
\begin{equation}
\lambda=\frac{C_F\alpha_s}{2\sqrt{-\frac{E}{m_t}}}.
\end{equation}

The Feynman rules required for higher-order computations have been
derived in~\cite{Beneke:2013PartI}. As also discussed there 
the soft matching of the two-point
function is trivial and hence 
\begin{equation}
  \Pi^{(a)}(q^2) = \frac{N_c c_a^2}{2m_t^4}\,\frac{d-2}{d-1}\, G^P(E),
 \label{eq:PiaPNRQCD}
\end{equation}
where now 
\begin{equation}
 \begin{aligned}
  G^P(E)=&\hspace*{0.15cm}\frac{i}{8N_c(d-2)}\int d^dx\,
e^{iEx^0}\Braket{0|T\left([\psi^\dagger\Gamma^i\chi]^\dagger(x)[\psi^\dagger\Gamma^i\chi](0)\right)|0}_{\text{PNRQCD}}\\
  =&\hspace*{0.15cm}\frac{i}{2N_c}\int d^dx\,
e^{iEx^0}\Braket{0|T\left([\chi^\dagger i D^i\psi](x)[\psi^\dagger i D^i\chi](0)\right)|0}_{\text{PNRQCD}}
 \end{aligned} 
\label{eq:GP}
\end{equation}
is the P-wave Green function at the origin. 
In passing to the second equation, we used that due to the
spin-independence of the Coulomb potential~\eqref{eq:VCoulomb} the
$(d-1)$-dimensional spin algebra in~\eqref{eq:GP} can be
evaluated once and for all. 
In PNRQCD perturbation theory up to NLO, the P-wave correlation
function reads
\begin{eqnarray}
    G^P(E) & = & G_0^P(E)+\delta_1G^P(E)+\dots \nonumber \\
    &=& \tilde{\mu}^{4\epsilon}
    \int\frac{d^{d-1}\mathbf{p}}{(2\pi)^{d-1}}
    \int\frac{d^{d-1}\mathbf{p}'}{(2\pi)^{d-1}}\,
    \mathbf{p}\cdot\mathbf{p}' \times
    \bigg[G_0(\mathbf{p},\mathbf{p}';E) \nonumber \\
    && + \,\tilde{\mu}^{4\epsilon}
    \int\frac{d^{d-1}\mathbf{p}_1}{(2\pi)^{d-1}}
    \int\frac{d^{d-1}\mathbf{p}_2}{(2\pi)^{d-1}}\,
    G_0(\mathbf{p},\mathbf{p}_1;E)\,
    i\left(-\frac{\alpha_s^2C_F}{(\mathbf{p}_1-\mathbf{p}_2)^2}\mathcal{V}_C^{(1)}\right)
    iG_0(\mathbf{p}_2,\mathbf{p}';E)\nonumber  \\
    && + \,\dots\bigg],
 \label{eq:GFPNRQCDexpansion}
\end{eqnarray}
where we have inserted the NLO Coulomb potential. As yet we have
neglected the sizable width of the top quark
$\Gamma_t = 1.33$~GeV. For the leading-order S-wave contribution the
width effect is accounted for by the replacement $E\ra E+i\Gamma_t$ in the
spectral function~\cite{Fadin:1987wz,Fadin:1988fn}. We follow this
prescription to {\em define} the pure QCD calculation of the 
pair-production cross section and therefore assume henceforth that 
$E$ can take complex values. Due to
the $\mathbf{p}\cdot\mathbf{p}'$ factor 
already the leading order result for the P-wave
contains ultraviolet divergences of the form $E/\epsilon$. Following
the above description yields poles proportional to $\Gamma_t/\epsilon$
in the imaginary part, and a scale-dependence related to these poles. 
To make this explicit, we distinguish between the scale $\mu_r$ at
which we evaluate the running coupling $\alpha_s=\alpha_s(\mu_r)$ and
the scale $\mu_w$ that arises from the finite-width divergences. 
While the residual $\mu_r$ scale-dependence must always be of 
higher-order, the
divergences and the dependence on the scale $\mu_w$ have to cancel
against non-resonant electroweak corrections involving a 
$Wb$-loop correction to the off-shell top propagator~\cite{Beneke:2008cr}. 
So far only the $1/\epsilon$ pole corresponding to the LO P-wave correlation 
function is known~\cite{Jantzen:2013gpa,Hoang:2004tg,Hoang:2010gu}. The finite
term is required to cancel the corresponding scheme dependence.  
Since the finite term as well as the NNNLO non-resonant terms related 
to the NLO   P-wave correlation 
function are presently unknown, we keep the dependence on $\mu_w$ 
(and the associated poles) explicit in our analytical result.

\section{Computation of the P-wave Green
  function\label{sec:computation}}

\subsection{Leading order\label{sec:LO}}

Simple power counting shows that ladder diagrams with up to four loops
are ultraviolet divergent, as compared to two loops for the S-wave, due to the
additional factor $\mathbf{p}\cdot\mathbf{p}'$ 
in~\eqref{eq:GFPNRQCDexpansion}. These diagrams, the sum of which is
denoted by $G_0^{P(\leq3ex)}$, therefore have to be computed in $d$
dimensions. We have used FIRE~\cite{Smirnov:2008iw,Smirnov:2013dia} to
perform the reduction to a small set of master integrals. Results for
these master integrals are available in the
literature~\cite{Schroder:2005va} and are in agreement with special
cases of the more general calculation for the NLO contribution presented below. 
We obtain
\begin{equation}
 G_0^{P(\leq3ex)}(E)=\frac{m_t^4C_F^3\alpha_s^3}{32\pi\lambda^3}
+4\pi C_F\alpha_s\left[I_{\text{P}}^{00}[1]+I_{\text{P}}^{10}[1]+I_{\text{P}}^{20}[1]\right],
\end{equation}
where we have given the contribution from the one-loop, zero-gluon exchange
diagram explicitly. It exhibits the characteristic 
$1/\lambda^3 \sim E^{3/2} \sim v^3$ threshold behaviour of P-wave 
production. The higher-loop integrals $I_{\text{P}}^{(n-1)0}[1]$, 
corresponding to diagrams with $n$ gluon exchanges, are given in 
App.~\ref{sec:diagrams}. (The notation is explained in more detail in 
the context of the NLO calculation after~\eqref{eq:IPa}.)
The remaining part $G_0^{P(\geq4ex)}$ is finite and can be
calculated in $d=4$ dimensions. We perform this calculation 
in position space. Eq.~(\ref{eq:GFPNRQCDexpansion}) implies 
\begin{equation}
\begin{aligned}
 G_0^{P(\geq4ex)}(E)&\hspace*{0.15cm} 
=\mathop{\text{lim}}\limits_{\mathbf{x},\mathbf{y}\ra0}\Braket{\nabla_x\cdot\nabla_{y}\; G_0^{(\geq4ex)}(\mathbf{x},\mathbf{y};E)}\\
 &\hspace*{0.15cm}
=\mathop{\text{lim}}\limits_{\mathbf{x},\mathbf{y}\ra0}\frac{1}{(4\pi)^2}\int d\Omega_{\mathbf{x}}\int d\Omega_{\mathbf{y}}\left[\nabla_x\cdot\nabla_{y}\; G_0^{(\geq4ex)}(\mathbf{x},\mathbf{y};E)\right],
\end{aligned}
 \label{eq:G0Pmorethan4ex}
\end{equation}
where $\Braket{\dots}$ denotes the angular mean of the respective
expression. This projects out the P-wave, $l=1$, component of the Green
function $G_0(\mathbf{r},\mathbf{r}';E)$. 
The expression~\eqref{eq:G0Pmorethan4ex} can be computed
using the representation~\eqref{eq:LOGFintegralrep} with appropriate
subtractions for the parts with less than three gluon exchanges. The
sum $G_0^{P(\leq3ex)}+G_0^{P(\geq4ex)}$ gives the expression for the
correlation function in dimensional regularization: 
\begin{equation}
  \begin{aligned}
    G_0^P(E) = \frac{m_t^4C_F^3\alpha_s^3}{32\pi\lambda^3}
    \bigg[& 1 -
      \left( \frac{1}{2\epsilon} + 2L_\lambda^w +4 \right) \lambda
      - 3\lambda^2
      + \left( \frac{1}{4\epsilon} + 2L_\lambda^w + \frac{7}{2}
      \right) \lambda^3 \\
      & + 2(\lambda-\lambda^3)\hat{\psi}(2-\lambda) \bigg],
  \end{aligned}
  \label{eq:GFMSbar}
\end{equation}
where
\begin{equation}
\hat{\psi}(z)=\gamma_E+\psi(z),
\hspace{1cm}
L_\lambda^a=\ln\left(\frac{\lambda\mu_a}{m_t\alpha_sC_F}\right)
=-\frac{1}{2}\ln\left(\frac{-4m_tE}{\mu_a^2}\right),
\end{equation}
with $a\in \{r,w\}$, $\gamma_E$ the Euler-Mascheroni constant, and $\psi(z)$ 
the logarithmic derivative of the Gamma function. 
In (\ref{eq:GFMSbar}) subtracting the 
$1/\epsilon$ poles gives the result in the $\overline{\text{MS}}$ scheme. 
Note that due to the overall $1/\lambda^3$ factor only the term 
proportional to $\lambda$ in square brackets results in a 
finite-width divergence. We have checked that this divergence in the  
imaginary part of the LO Green function~\eqref{eq:GFMSbar} agrees 
with the result given
in~\cite{Jantzen:2013gpa,Hoang:2001mm}. Neglecting the width of the
top, the imaginary part is finite, and reads 
\begin{eqnarray}
 \text{Im}\!\left[G_0^P(E)\right] &=&
\left(\frac{m_tC_F\alpha_s}{2}\right)^{\!5}
\sum\limits_{n=2}^\infty\frac{n^2-1}{n^5}\,\delta(E-E_n)
\nonumber\\
&& +\,\frac{m_t^4}{4\pi}\left(\frac{E}{m_t}
+\frac{C_F^2\alpha_s^2}{4}\right)
\frac{\pi C_F\alpha_s}{1-e^{-C_F\alpha_s\pi/v}}\,\theta(E),
 \label{eq:IMLO}
\end{eqnarray}
where $v\equiv \sqrt{E/m_t}$, and 
$E_n=-(m_tC_F^2\alpha_s^2)/(4n^2)$ with $n\geq2$ are the $l=1$
bound state energies. Eq.~(\ref{eq:IMLO}) agrees with~\cite{Bigi:1991mi}.

\subsection{Next-to-leading order\label{sec:NLO}}

\begin{figure}[t]
\centering
 \includegraphics[width=0.85\textwidth]{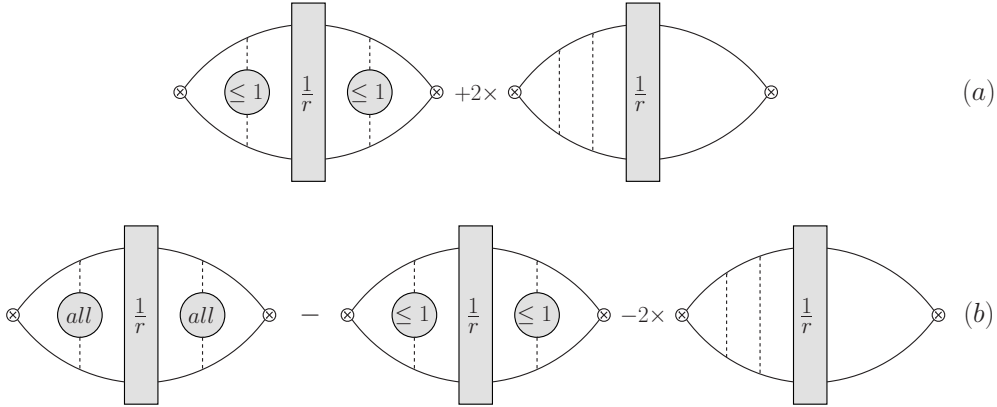}
 \caption{We split the NLO correction to the Green function into a
   divergent part (a) that contains all diagrams with up to four loops
   and the finite remainder (b).\label{fig:SingleCoulombInsertionab}}
\end{figure}

Analogous to the S-wave computation \cite{Beneke:2013PartII}, 
we define the single-insertion function
\begin{equation}
 I_{\text{P}}[x+u]=
 \int\left[\prod\limits_{i=1}^4\frac{d^{d-1}\mathbf{p}_i}{(2\pi)^{d-1}}\right]
 \mathbf{p}_1\cdot\mathbf{p}_4\;
 \tilde{G}_0(\mathbf{p}_1,\mathbf{p}_2;E)\frac{1}{(\mathbf{q}_{23}^2)^x}
 \left(\frac{\mu^2}{\mathbf{q}_{23}^2}\right)^{\!u}
 \tilde{G}_0(\mathbf{p}_3,\mathbf{p}_4;E),
\label{eq:IP}
\end{equation}
where $\mathbf{q}_{ij}=\mathbf{p}_i-\mathbf{p}_j$.
In terms of this the NLO correction to the Green function contained in 
(\ref{eq:GFPNRQCDexpansion}) is given by
\begin{equation}
 \delta_1G^P(E)=C_F\alpha_s^2
 \left[\frac{\beta_0}{\epsilon}\Big(I_{\text{P}}[1+\epsilon]-
 I_{\text{P}}[1]\Big)+a_1(\epsilon)I_{\text{P}}[1+\epsilon]\right].
\label{eq:delta1G}
\end{equation}
As for the LO Green function divergences only occur in diagrams with
up to four loops. We therefore split the NLO correction into a
divergent (a) and a finite part (b) as indicated in
Fig.~\ref{fig:SingleCoulombInsertionab}. Since the top quark width
$\Gamma_t$ cannot be neglected, the imaginary part of $(a)$ contains
divergences of the type $\Gamma_t/\epsilon$ arising from poles of the
form $E/\epsilon$. Thus, contrary to the S-wave case, where no such 
divergences are present in the computation of corrections from the 
Coulomb potential, the NLO Coulomb
potential~\eqref{eq:VCoulombcoefficients} cannot be expanded in
$\epsilon$ prior to integration 
in the computation of the loop integrals of part $(a)$. 
However, the momentum integrals of part $(b)$ are
finite and the potential can be expanded before integration here.

To deal with part (b), we start with the computation of the 
complete $I_{\text{P}}[1+u]$ in position space and later perform the 
necessary subtractions to obtain $I_{\text{P}}^{(b)}[1+u]$: 
\begin{eqnarray}
 I_{\text{P}}[1+u] &=&
    \mathop{\text{lim}}\limits_{\mathbf{x},\mathbf{y}\ra0}
    \Braket{ \left(\nabla_x\cdot\nabla_y\right)
      \int d^{d-1}\mathbf{r}\,G_0(\mathbf{x},\mathbf{r};E)
      \frac{\mu^{2u}\left(\mathbf{r}^2\right)^{u-\frac12}}
      {4\pi\Gamma(1+2u)\cos(\pi u)}
      G_0(\mathbf{r},\mathbf{y};E)} \nonumber \\
    && \hspace*{-2cm} = \,\mathop{\text{lim}}\limits_{\mathbf{x},\mathbf{y}\ra0}
    \frac{9m_t^2(2p)^6\mu^{2u}}
    {(4\pi)^3\Gamma(1+2u)\cos(\pi
      u)\,\Gamma(2+\lambda)^2\Gamma(2-\lambda)^2}\nonumber \\
    && \hspace*{-1.8cm} \times \,\int\limits_0^1dt_1\int\limits_0^1dt_2
    \left[(1-t_1)(1-t_2)\right]^{1+\lambda}
    \left[t_1t_2\right]^{1-\lambda}
    \int\limits_1^\infty ds_1\int\limits_1^\infty ds_2
    \left[s_1s_2\right]^{1+\lambda}\left[(s_1-1)(s_2-1)\right]^{1-\lambda}
\nonumber \\
    && \hspace*{-1.8cm} \times \, \Braket{ \left(\nabla_x\cdot\nabla_y\right)
      \int d^{d-1}\mathbf{r} \left(\mathbf{x}\cdot\mathbf{r}\right)
      \left(\mathbf{y}\cdot\mathbf{r}\right)
      \left(\mathbf{r}^2\right)^{u-\frac12}
      e^{-p\left[x(1-2t_1)+y(1-2t_2)+2r(s_1+s_2-1)\right]}}.
 \end{eqnarray}
We proceed by first solving the integral over $\mathbf{r}$:
\begin{equation}
  \int d^{d-1}\mathbf{r}\ r^ir^j\,r^{2u-1} e^{-2pr(s_1+s_2-1)}=
  \frac{\delta^{ij}}{d-1}\,
  \frac{2\pi^{(d-1)/2} \Gamma(d+2u)}{\Gamma((d-1)/2) }
   \left[2p(s_1+s_2-1)\right]^{-d-2u}.
\end{equation}
Taking the derivatives and the limit $\mathbf{x},\mathbf{y}\ra0$, we
can perform the integrations over $t_1$ and $t_2$ and obtain 
\begin{equation}
  \begin{aligned}
    I_{\text{P}}[1+u] = &\hspace*{0.15cm}
    \frac{m_t^2(2p)^{6-d-2u}\mu^{2u}}{4(4\pi)^3\Gamma(1+2u)\cos(\pi u)}
    \frac{2\pi^{(d-1)/2}\Gamma(d+2u)}{\Gamma((d-1)/2)} \\
    & \times \int\limits_1^\infty ds_1\int\limits_1^\infty ds_2\,
    \frac{\left[s_1s_2\right]^{1+\lambda}
      \left[(s_1-1)(s_2-1)\right]^{1-\lambda}}{(s_1+s_2-1)^{d+2u}}.
  \end{aligned}
\end{equation}
After the necessary subtraction the remaining part $(b)$ is finite and
can be computed in $d=4$ dimensions. We obtain
\begin{equation}
 I_{\text{P}}^{(b)}[1+u]=\frac{m_t^2p^2}{(4\pi)^2}
\left(-\frac{\mu^2}{4m_tE}\right)^uj_{\text{P}}(u),
 \label{eq:Ib}
\end{equation}
where 
\begin{eqnarray}
 j_{\text{P}}(u)&=&\frac{\Gamma(4+2u)}{\Gamma(1+2u)\cos(\pi u)}
\int\limits_0^\infty dt_1\int\limits_0^\infty dt_2\frac{t_1t_2(1+t_1)(1+t_2)}{(1+t_1+t_2)^{4+2u}}\Bigg[\left(\frac{(1+t_1)(1+t_2)}{t_1t_2}\right)^\lambda
\nonumber\\
&&-\,1-\lambda\log\left(\frac{(1+t_1)(1+t_2)}{t_1t_2}\right)
-\frac{\lambda^2}{2}\log^2\left(\frac{(1+t_1)(1+t_2)}{t_1t_2}\right)\Bigg].
 \label{eq:ju}
\end{eqnarray}
Here the last three terms are the first three terms in the expansion 
in $\alpha_s$, which corresponds to up to two gluon exchanges to the 
left or right of the NLO Coulomb potential insertion. This is precisely 
part (a), hence the above, subtracted expression is part (b) as desired. 
A strategy to solve this kind of integral is presented 
in~\cite{Beneke:2013PartII}. We obtain
\begin{eqnarray}
 j_{\text{P}}(0) &=&
-1+\left(\frac{\pi^2}{3}-2\right)\lambda+3\zeta (3)\lambda ^2+\left(1-3\lambda^2\right)\hat{\psi}(2-\lambda)+\left(\lambda^3-\lambda\right)\psi_1(2-\lambda),
\nonumber\\[-0.4cm]
&& \label{eq:j0}
\\
  j'_{\text{P}}(0)&=&\frac{\pi ^2}{6}-\frac{50}{9}+\left[-4+\frac{2 \pi ^2}{3}-4 \zeta (3)\right]\lambda+\left[\frac{34}{3}+\frac{\pi ^2}{6}-\frac{\pi ^4}{180}+6 \zeta (3)\right]\lambda ^2 \nonumber\\
  &&+\,\left[4+6\lambda-10\lambda^2\right]\hat{\psi}(2-\lambda)+\left(3 \lambda ^2-1\right)\left[\hat{\psi}(2-\lambda)^2-3\psi_1(2-\lambda)\right] \nonumber\\
  &&+\,\left(\lambda^3-\lambda\right)\left[\left(\frac{22}{3}-2\hat{\psi}(2-\lambda)\right)\psi_1(2-\lambda)+\psi_2(2-\lambda)\right] \nonumber\\
  &&+\,\frac{3}{2(\lambda -2)}\, _4F_3(1,1,4,4;5,5,3-\lambda;1).
 \label{eq:jprime0}
\end{eqnarray}
Here $\psi_n(z)$ is the $n$th derivative of the psi function. 
We provide some useful formulas for the evaluation of the generalized
hypergeometric function $_4F_3$ in App.~\ref{sec:pFq}. In terms of
$j_{\text{P}}(u)$ and $j'_{\text{P}}(u)$ from~\eqref{eq:j0}
and~\eqref{eq:jprime0} we can write part $(b)$ of~\eqref{eq:delta1G} as
\begin{equation}
 \begin{aligned}
 \delta_1G^{P(b)}(E)=&\hspace*{0.15cm}
C_F\alpha_s^2\left[\beta_0\frac{d}{du}I_{\text{P}}^{(b)}[1+u]\Big|_{u=0}
+a_1(0)I_{\text{P}}^{(b)}[1]\right]\\
 =&\hspace*{0.15cm} 
\frac{m_t^4\alpha_s^4C_F^3}{64\pi^2\lambda^2}
\left[\beta_0\left(2j_{\text{P}}(0)L_\lambda+j_{\text{P}}'(0)\right)
+a_1(0)j_{\text{P}}(0)\right],
 \end{aligned}
 \label{eq:delta1GPb}
\end{equation}
where we have expanded the NLO Coulomb potential 
(\ref{eq:VCoulombcoefficients}) and written the
logarithm of $\mathbf{q}^2$ as a derivative at zero $u$.

The divergent part $(a)$ is given by
\begin{equation}
 I_{\text{P}}^{(a)}[1+u]=I_{\text{P}}^{00}[1+u]+2I_{\text{P}}^{10}[1+u]
+2I_{\text{P}}^{20}[1+u]+I_{\text{P}}^{11}[1+u],
 \label{eq:IPa}
\end{equation}
where the $I_{\text{P}}^{nm}[1+u]$ denotes the contribution to the
single insertion function from the diagram with $n$ potential gluon
exchanges to the left and $m$ to the right of the potential
insertion. Part $(a)$ then takes the form of~\eqref{eq:delta1G} 
with $I_P$ replaced by~\eqref{eq:IPa}, and requires the calculation 
of some four-loop diagrams in dimensional regularization. 
The results for the  $I_{\text{P}}^{nm}[1+u]$ needed are given in 
App.~\ref{sec:diagrams}. The complete NLO correction to the Green function 
in dimensional regularization given by the sum of parts $(a)$ and $(b)$ reads:
\begin{eqnarray}
  \lefteqn{\delta_1G^P(E)=-\frac{m_t^4\alpha_s^4C_F^3}{64\pi^2\lambda^2}}
  && \nonumber \\
  && \times \Bigg\{ \beta_0 \Bigg[
  \left(-\frac{1}{12 \epsilon ^2}+\frac{59}{9}+\frac{5 \pi ^2}{72}
    +4L_\lambda^r+2L_\lambda^rL_\lambda^w-\left(L_\lambda^w\right)^2\right)
  +\left(9+6L_\lambda^r\right)\lambda \nonumber \\
  && \phantom{\times\{ \beta_0}
  +\left(\frac{3}{40 \epsilon ^2}+\frac{1}{20 \epsilon }
    -\frac{344}{15}-\frac{\pi ^2}{8}
    -\frac{21}{2}L_\lambda^r+\frac{1}{2}L_\lambda^w-6L_\lambda^wL_\lambda^r
    +3\left(L_\lambda^w\right)^2\right)\lambda ^2 \nonumber \\
  && \phantom{\times\{ \beta_0}
  +\left[-4-6
    \lambda+10\lambda^2+2(3\lambda^2-1)L_\lambda^r\right]\hat\psi(2-\lambda)
  \nonumber \\
  && \phantom{\times\{ \beta_0}
  +\left(\lambda -\lambda ^3\right)
  \left[\psi_1(2-\lambda)\left(\frac{22}{3}+2L_\lambda^r-2\hat\psi(2-\lambda)\right)
    +\psi_2(2-\lambda )\right] \nonumber \\
  && \phantom{\times\{ \beta_0}
  +\left(3 \lambda ^2-1\right) \left[3 \psi_1(2-\lambda)-\hat\psi(2-\lambda)^2\right]
  +\frac{3}{4-2 \lambda } \, _4F_3(1,1,4,4;5,5,3-\lambda ;1)\Bigg]
  \nonumber \\
  && \phantom{\times}
  + a_1(\epsilon) \bigg[
  \frac{1}{6\epsilon} + 2+L_\lambda^w + 3\lambda
  - \left(\frac{3}{10 \epsilon} + \frac{26}{5} + 3L_\lambda^w \right) \lambda^2
  + \left(3 \lambda ^2 - 1\right) \hat\psi(2-\lambda) \nonumber \\
  && \phantom{\times + a_1(\epsilon)}
  +\left(\lambda -\lambda ^3\right) \psi_1(2-\lambda )\bigg]\Bigg\}.
\label{eq:delta1GP}
\end{eqnarray}
The dependence on the two scales $\mu_r$ and $\mu_w$ has been obtained
with the procedure described in App.~\ref{sec:diagrams}. Alternatively, 
the dependence on $\mu_r$ can be obtained using one-loop running 
of $\alpha_s$ in the LO Green function. The remaining logarithms of 
$\mu$ must then be assigned to $\mu_w$. We note that the dependence on
the scale $\mu_w$ is polynomial in $E$ and cancels in the 
imaginary part for $\Gamma_t=0$, 
which provides a useful consistency check.

\subsection{Pole resummation\label{sec:poles}}

At negative energies the P-wave Green function contains poles
corresponding to quark-antiquark bound states with angular momentum
$l=1$. In~\eqref{eq:delta1GP} they appear as poles of
polygamma and hypergeometric functions for positive integer
$\lambda\geq2$. Near these bound states the exact Green function has
the form
\begin{equation}
 G^P(E)\mathop{=}\limits^{E\ra E_n^P}\frac{|\psi'_n(0)|^2}
{E_n^P-E-i\epsilon}+\text{regular},
 \label{eq:GFpolesexact}
\end{equation}
where $E_n^P$ is the energy of the $n$th P-wave bound state and
$\psi'_n(0)$ the derivative of the corresponding wave function at the
origin. Both take on a perturbative expansion in the strong coupling
constant
\begin{equation}
 E_n^P=E_n^{P(0)}\left(1+\frac{\alpha_s}{4\pi}\,e_1^P
+\mathcal{O}(\alpha_s^2)\right),
\hspace{0.5cm}|\psi'_n(0)|^2=|\psi_n'^{(0)}(0)|^2
\left(1+\frac{\alpha_s}{4\pi}f_1^P+\mathcal{O}(\alpha_s^2)\right).
\end{equation}
The NLO Green function expanded in $\alpha_s$ therefore takes the form
\begin{eqnarray}
    G^P(E)&\mathop{=}\limits^{E\ra E_n^{P(0)}}&
    \frac{|\psi_n'^{(0)}(0)|^2}{E_n^{P(0)}-E-i\epsilon}
    +\frac{\alpha_s}{4\pi}
    \left(\frac{f_1^P|\psi_n'^{(0)}(0)|^2}{E_n^{P(0)}-E-i\epsilon}
      -\frac{e_1^P|\psi_n'^{(0)}(0)|^2E_n^{P(0)}}{\left(E_n^{P(0)}-E-i\epsilon\right)^2}\right)\nonumber \\
    &&+\,\mathcal{O}(\alpha_s^2)+\text{regular}.
  \label{eq:GFpolesexpanded}
\end{eqnarray}
The singular terms near $E_n^{P(0)}$ can be resummed into a single 
pole to all orders by
subtracting~\eqref{eq:GFpolesexpanded} from the Green function and
adding~\eqref{eq:GFpolesexact} with the energies and derivatives of
the wave function in NLO approximation \cite{Beneke:1999qg}. 
The leading-order expressions
can be read off from the imaginary part~\eqref{eq:IMLO} of the LO
result:
\begin{equation}
 E_n^{P(0)}=-\frac{m_tC_F^2\alpha_s^2}{4n^2},\hspace{1cm}
|\psi_n'^{(0)}(0)|^2=\frac1\pi\left(\frac{m_tC_F\alpha_s}{2}\right)^{\!
  5}\,
\frac{n^2-1}{n^5}.
\end{equation}
To obtain the NLO corrections we expand~\eqref{eq:delta1GP} for
$\lambda$ near positive integer $n$. We find 
\begin{eqnarray}
  \lefteqn{\delta_1G^P(E)\,
    \mathop{=}\limits^{\lambda\ra n}\,
    \frac{m_t^4 \alpha_s^4C_F^3}{4(4\pi)^2}} && \nonumber \\
  && \times\Bigg\{
  \frac{n^2-1}{n^2(n-\lambda )}
  \left[2\beta_0\left(2L_n^r+4+\frac{3}{n-1}-\frac{4n^2}{n^2-1}\hat\psi(n+2)
      -2n\psi_1(n+2)\right)+2a_1\right] \nonumber \\
  && \phantom{\times}
  +\frac{n^2-1}{n(n-\lambda)^2}\left[2\beta_0(L_n^r+\hat\psi(n+2))+a_1\right]
  +\text{regular}\Bigg\}.
\end{eqnarray}
With this it is straightforward to obtain:
\begin{align}
 e_1^P=&2a_1+4\beta_0\left[L_n^r+\hat{\psi}(n+2)\right],\\
 f_1^P=&5a_1+2\beta_0\left[5L_n^r+4+\frac{3}{n-1}-\frac{n^2+3}{n^2-1}\hat{\psi}(n+2)-2n\psi_1(2+n)\right],
\end{align}
where $L_n^r=\ln\left(n\mu_r/(m_tC_F\alpha_s)\right).$ We have checked
that this agrees with the results
of~\cite{Penin:1998ik,Penin:1998mx}.

\section{P-wave top-pair production cross
  section\label{sec:crosssection}}

In this section we discuss the phenomenological aspects of the P-wave
contribution to the top-pair production cross section near
threshold. All expressions so far have employed the pole mass
definition of the top quark. Since the pole mass suffers from an
infrared renormalon
ambiguity~\cite{Beneke:1994sw,Bigi:1994em,Beneke:1994rs}, which is 
not present in the top-pair cross section itself, we
show results using the PS mass definition~\cite{Beneke:1998rk}, 
which eliminates this spurious infrared sensitivity. This
has been implemented in the PS Shift (PSS) and PS Insertion (PSI)
schemes~\cite{Beneke:2013PartII}. Denoting by $\delta m_t$ 
the difference between the pole mass and the PS mass, 
the former is defined by
\begin{equation}
  G^{\text{PSS}}(\sqrt{s},m_t^{\text{PS}}) =
  G^{\text{pole}}(\sqrt{s}, m_t^{\text{PS}} + \delta m_t),
\end{equation}
where the value of $\delta m_t$ is order-dependent. That is, in LO, 
we only use the leading-order expression $\delta m_t\propto \mu_f\alpha_s$, 
whereas in NLO,  $\delta m_t$ includes the $\mu_f\alpha_s^2$ term that contains
the $a_1$ correction to the Coulomb potential~\cite{Beneke:1998rk}.\footnote{
However, when our result is combined with the NNNLO S-wave
contribution, $\delta m_t$ should be used at NNNLO as well.} The 
PSI scheme is obtained by re-expanding the right-hand
side in $\alpha_s$. We find, however, that the difference between the 
two schemes is very small and would not be visible
in the figures below. We therefore only show the results in the 
PSS scheme. For the numerics we adopt the parameter values 
$\alpha_s(M_Z) = 0.1184$, $m_t^{\text{PS}} 
\equiv m_t^{\text{PS}}(\mu_f=20\,\mbox{GeV}) = 171\,\mbox{GeV}$, 
which corresponds to a LO (NLO) pole mass of $m_t = 172.025
\,(172.433)\,$GeV, the top-quark width $\Gamma_t = 1.33\,$GeV and the
Weinberg  
angle $\sin^2 \theta_w = 0.23$. The strong coupling is evolved to the 
scale $\mu_r$ in the four-loop approximation. 

Since $G^P(E)$ is divergent there exists an ambiguity in (\ref{eq:PiaPNRQCD}) 
whether $G^P(E)$ or $\Pi^{(a)}(q^2)$ should be minimally subtracted, 
since the factor that relates both depends on $d$.  
This ambiguity is only resolved when the finite term of the non-resonant 
contribution is computed. Since the non-resonant calculation does not 
refer to any kind of non-relativistic approximation it is natural to 
define the non-resonant and resonant contribution by minimal subtraction 
of $\Pi^{(a)}(q^2)$. This will be assumed in the following. To be 
precise, we write the expansion of the hard coefficient (\ref{eq:ca})
in the form $c_a = 1+ [c_a^{(1)} +\epsilon c_a^{(1\epsilon)}]+  
\mathcal{O}(\alpha_s^2)$, and the LO and NLO 
Green functions (\ref{eq:GFMSbar}), (\ref{eq:delta1GP}), respectively, as 
\begin{equation}
G_0^P(E) = \frac{1}{\epsilon} 
\,G_{0, \rm div}+ G_{0,\overline{\mbox{\tiny MS}}}^P,
\qquad
\delta_1 G^P(E) =
\frac{1}{\epsilon^2} \,\delta_1 G_{\rm div2}+ 
\frac{1}{\epsilon} \,\delta_1 G_{\rm div1} +
 \delta_1 G_{\overline{\mbox{\tiny MS}}}^P.
\end{equation} 
Then the expansion of $\Pi^{(a)}(q^2)$ to NLO in minimal subtraction 
is given by 
$\Pi^{(a)}(q^2) = \Pi_0^{(a)}(q^2) + \delta_1\Pi^{(a)}(q^2)$ with 
\begin{eqnarray}
\text{Im}\,[\Pi_0^{(a)}(q^2)] &=& \frac{N_c}{2m_t^4}\,
\frac{2}{3}\, \text{Im}\,[G_{0,\overline{\mbox{\tiny MS}}}^P(E)] - 
\frac{2}{9}\, \frac{N_c}{32\pi}\frac{\alpha_s C_F\Gamma_t}{m_t},
\label{PiMSLO}\\
\text{Im}\,[\delta_1 \Pi^{(a)}(q^2)] &=& 
\frac{N_c}{2m_t^4}\,
\frac{2}{3}\, \text{Im}\,[\delta_1 G_{\overline{\mbox{\tiny MS}}}^P(E)] + 
2 c_a^{(1)} \,\text{Im}\,[\Pi_0^{(a)}(q^2)]
\nonumber\\ 
&&  + 2 c_a^{(1\epsilon)} \,\frac{2}{3}\, 
\frac{N_c}{32\pi}\frac{\alpha_s C_F\Gamma_t}{m_t}
+ \left(\frac{\beta_0}{81} - \frac{a_1(0)}{27}\right)
 \frac{N_c}{32\pi^2} \frac{\alpha^2_s C_F\Gamma_t}{m_t},
\label{PiMSNLO}
\end{eqnarray} 
where the constant terms proportional to $\Gamma_t/m_t$ 
arise from the divergent 
parts of $G^P(E)$ multiplying the order $\epsilon$ and $\epsilon^2$ terms 
of the factor $(d-2)/(d-1)\times c_a^2$, which  
relates  $G^P(E)$ to $\Pi^{(a)}(q^2)$.\footnote{The scale $\mu$ 
in the logarithm in $c_a^{(1\epsilon)}$, see (\ref{eq:ca}), should 
be identified with $\mu_w$.} Note that we assume here that $\Gamma_t$ 
takes its numerical, four-dimensional, physical value, while for 
an analytic combination with the (yet unknown) non-resonant 
cross section, one eventually needs to use the analytic, $d$ dimensional, 
leading-order expression for the top-quark width. The constant terms 
in (\ref{PiMSLO}) and (\ref{PiMSNLO}) shift the 
cross section only by a tiny amount, so that the scheme-dependence related 
to the resonant-nonresonant separation that 
can be resolved only once the non-resonant cross section is fully known, 
is not relevant for the following discussion.

\begin{figure}[t]
  \centering
  \includegraphics[width=0.7\textwidth]{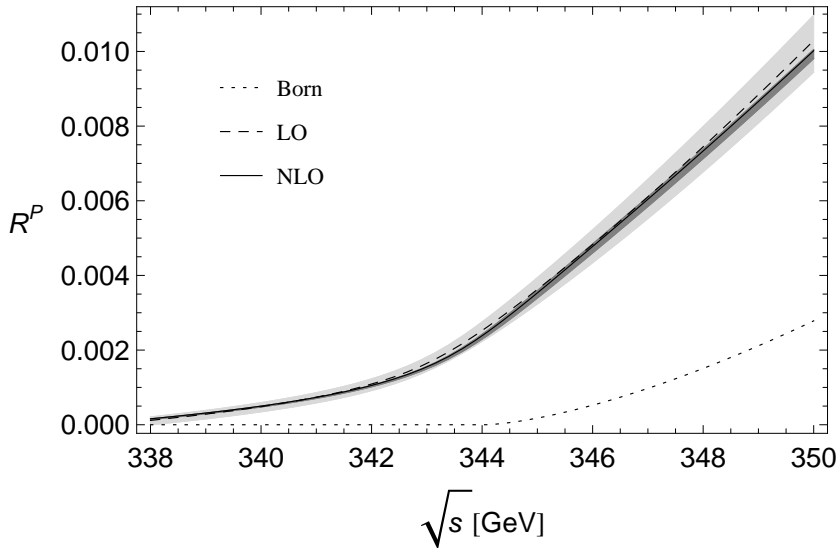}
  \caption{The P-wave contribution to the R ratio in the PSS scheme
   for $m_t^{\text{PS}}=171$~GeV, $\Gamma_t=1.33$~GeV, 
   $\mu_r=80$~GeV, and $\mu_w=m_t^{\text{PS}}$. The dashed and solid
   lines denote the LO and NLO contributions, respectively. The shaded
   regions show the respective scale uncertainties for a variation of
   $\mu_r$ in the region $[50\text{ GeV},m_t^{\text{PS}}]$. The dark-shaded 
   band for the small NLO scale variation is hardly visible. The
   dotted line denotes the Born-level result.\label{fig:Ra}}
\end{figure}

In Fig.~\ref{fig:Ra} we show the LO and NLO P-wave contributions to the 
$R$ ratio in the PSS scheme, employing the expressions (\ref{PiMSLO}), 
(\ref{PiMSNLO}), 
and including pole resummation up to the $n=6$ bound-state pole.  
The overall size relative to the S-wave is below $1\%$
in the threshold region, because in addition to the $v^2$ suppression
the ratio of the couplings $(v_e^2+a_e^2)a_t^2/e_t^2\approx0.28$ is
small.\footnote{The possibility to extract the P-wave contribution
  using different beam polarizations was discussed in~\cite{Kuhn:1999hw}.} 
In order to determine the theoretical 
uncertainty we vary the renormalization scale $\mu_r$ in the range 
$[50\text{GeV},m_t^{\text{PS}}]$, while keeping $\mu_w=m_t^{\text{PS}}$
fixed. The latter scale is chosen of order of the hard scale 
in order to capture the logarithmically enhanced contribution of 
the unknown non-resonant part. We observe that the dependence on 
$\mu_r$ is much reduced at NLO, which implies
that perturbation theory works very well for the P-wave contribution. 
A comparison with the Born level result (dotted curve) shows a 
sizable difference and thus the importance of the Coulomb resummation, 
which enhances the cross section. Note that unlike the S-wave case, 
the strong Coulomb attraction does not lead to a peak structure in 
the energy-dependence of the cross section, since the residue of 
the lowest $n=2$ P-wave bound state is already too small to be visible given 
the large top-quark width. 

\begin{figure}[t]
  \centering
  \includegraphics[width=0.45\textwidth]{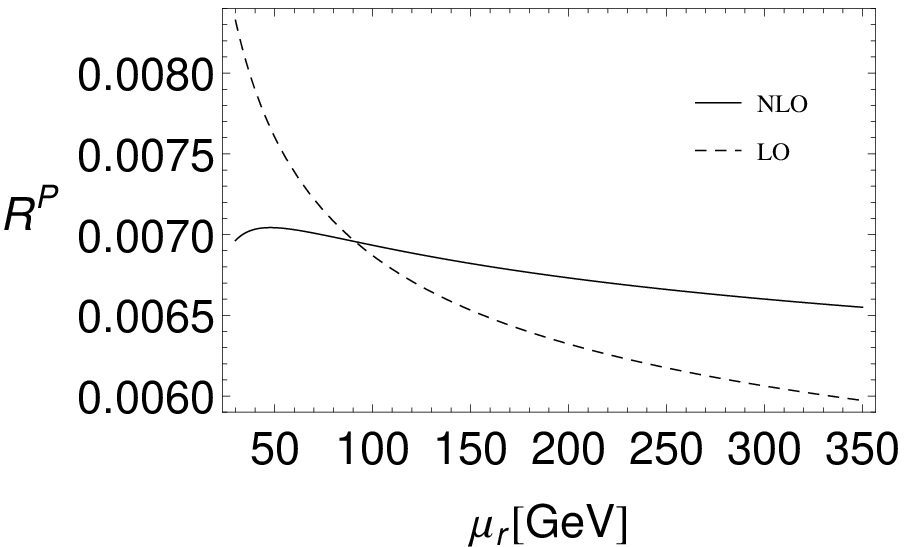}
  \includegraphics[width=0.45\textwidth]{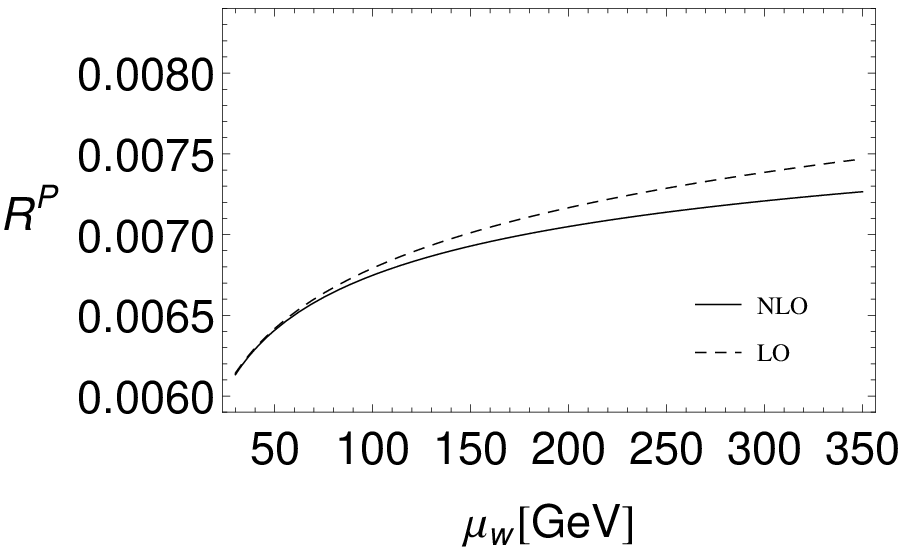}\\[1.5ex]
  \includegraphics[width=0.45\textwidth]{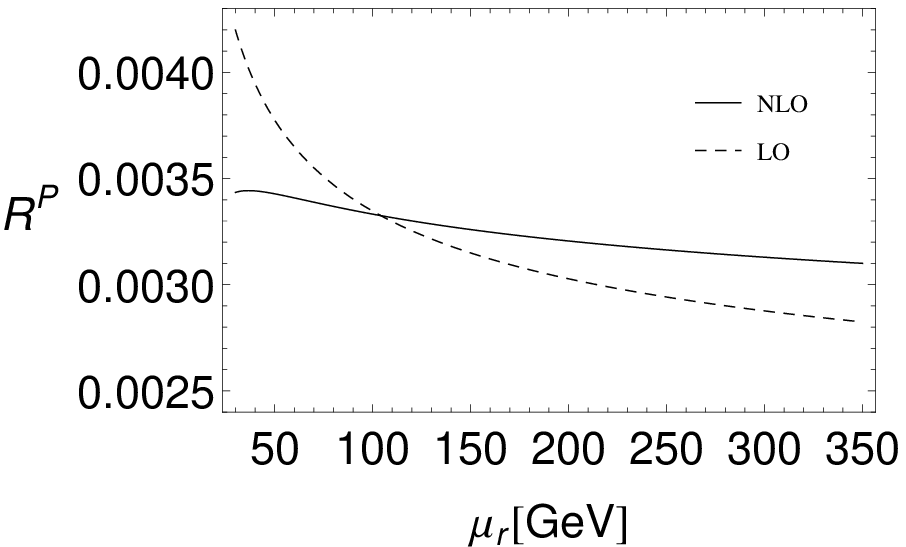}
  \includegraphics[width=0.45\textwidth]{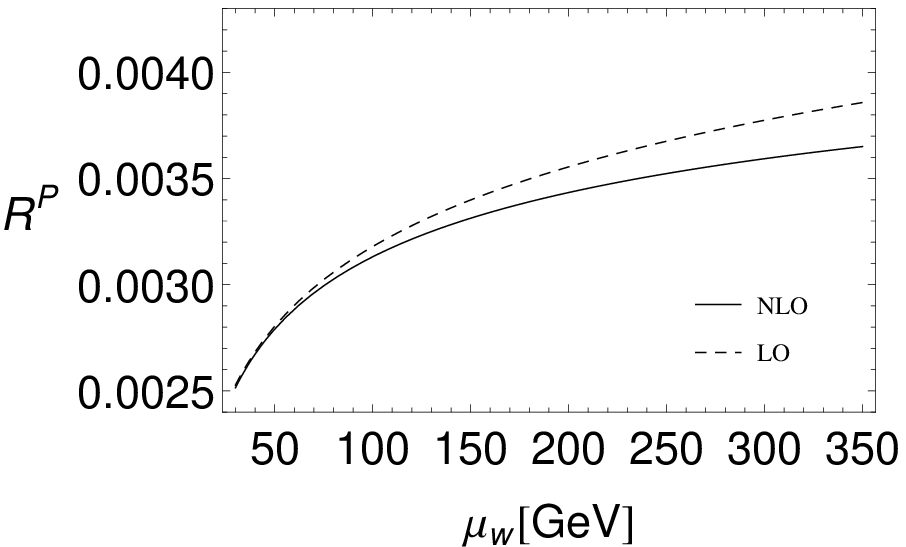}\\[1ex]
  \includegraphics[width=0.45\textwidth]{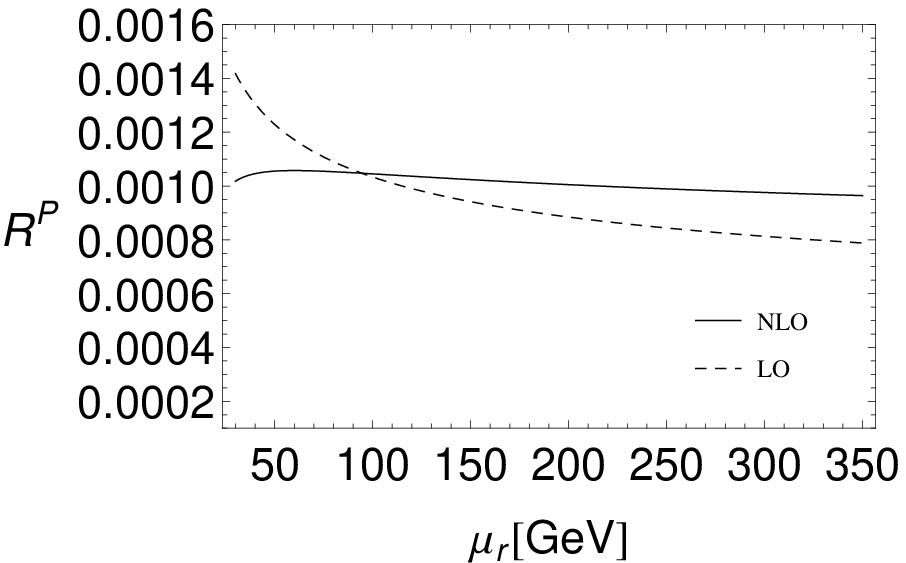}
  \includegraphics[width=0.45\textwidth]{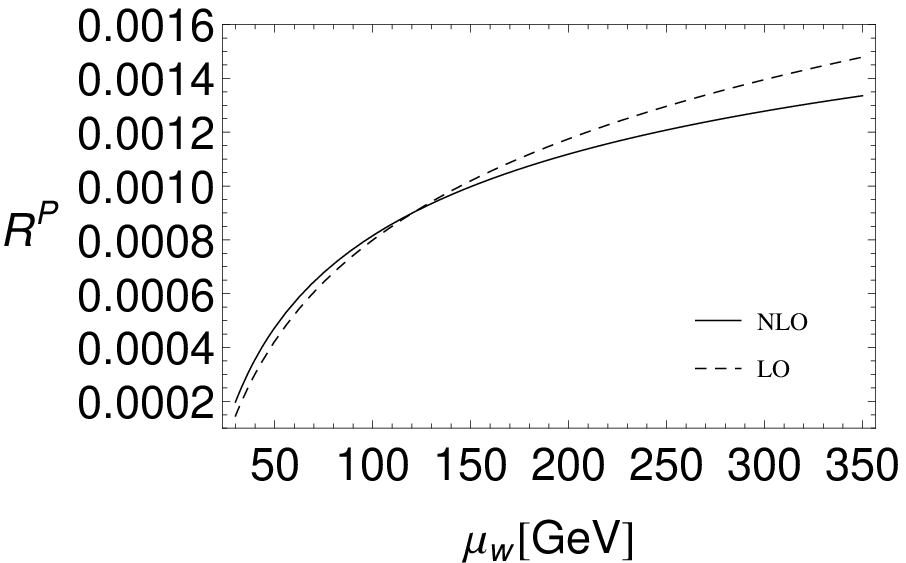}
  \caption{\label{fig::scales}The scale dependence of the P-wave
    contribution to the R ratio. The left and right plots show the
    dependence on $\mu_r$ for fixed $\mu_w$ and vice versa. The
    dependence is shown above (upper plots), close to (middle plots), and
    below  threshold (lower plots) corresponding to 
   $\sqrt{s} = 347.733, 344.866, 342\,$GeV, respectively. In the left plots 
the finite width scale is fixed to $\mu_w = m_t^{\text{PS}}$. In the 
right plots the coupling-renormalization scale is fixed to $\mu_r = 
80\,$GeV.}
\end{figure}

A comparison of the dependence on the two scales is shown in
Fig.~\ref{fig::scales} for three values of the energy (above, at and 
below threshold from top to bottom). We again see
that the dependence on the renormalization scale $\mu_r$ is strongly reduced at
NLO. This is not the case for the dependence on $\mu_w$,
which remains almost the same as at LO. This is expected, since 
the finite-width scale dependence does not cancel by performing higher-order 
QCD calculations. Rather, it has to cancel only when the non-resonant
corrections are added to the result. The fact that the finite-width 
scale dependence is dominant at NLO shows the importance of this
cancellation. In particular, the lower-right plot shows that the 
finite-width scale dependence changes the cross section by a large 
factor below threshold, precisely where the non-resonant contributions are 
expected to be important.

\section{Top-pair production in photon collisions\label{sec:photon}}

The photon collider option via the back-scattering
method~\cite{Ginzburg:1982yr} was studied in the Technical Design
Report for TESLA~\cite{Badelek:2001xb} and is also considered at the
ILC (see Sec.~12.6 of~\cite{Adolphsen:2013kya}). We discuss here only
$\gamma\gamma\to t\bar t$ collisions 
with opposite photon helicities, where the top pair is
produced in a P-wave state. We define the inclusive normalized cross section
$R_\gamma^{+-}=\sigma_{\gamma\gamma\ra t\bar{t}X}^{+-}/\sigma_0$. Near
the production threshold the cross section takes the
form~\cite{Penin:1998ik,Penin:1998mx} 
\begin{equation}
 R_\gamma^{+-}=\frac{32\pi N_ce_t^4}{m_t^4}\, 
C_h^{+-}(\alpha_s)\, 
\text{Im}\left[G^P(E)\right],
\end{equation}
with the hard matching coefficient
\begin{equation}
 C_h^{+-}(\alpha_s)=1-16C_F\frac{\alpha_s}{4\pi}.
\end{equation}
In the absence of a $d$ dimensional calculation of $C_h^{+-}$ we define 
here the R ratio, to which non-resonant contributions 
should eventually be added, by minimal subtraction of the P-wave 
Green function $G^P(E)$. 

\begin{figure}[t]
  \centering
  \includegraphics[width=0.7\textwidth]{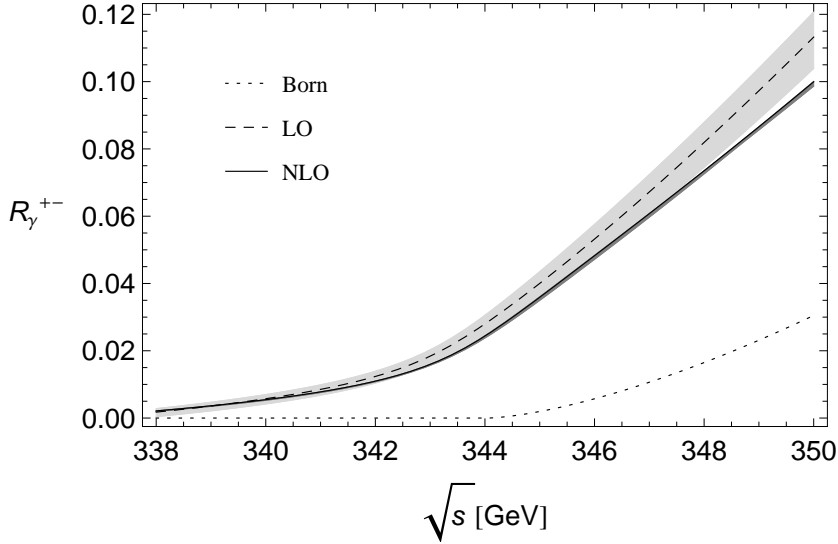}
  \caption{$R_\gamma^{+-}=\sigma_{\gamma\gamma\ra
      X\bar{t}t}^{+-}/\sigma_0$ at LO (dashed) and NLO (solid) in
   the PSS scheme. The same coding as in Fig.~\ref{fig:Ra} is
   adopted.\label{fig:R_pm}}
\end{figure}

We show results within the PSS scheme in Fig.~\ref{fig:R_pm}. The P-wave 
cross section is an order of magnitude larger than in $e^+e^-$
collisions, due the different size of the electroweak couplings and 
numerical prefactors. It can
be observed independently of the S-wave by adjusting the beam
polarizations. Since the $\gamma\gamma$ induced cross section differs 
from the $e^+e^-$ one only by the hard-matching coefficient and 
overall electroweak couplings, we observe essentially the same 
features as in the previous section. Most importantly, the theoretical 
uncertainty of the QCD contributions as measured by the residual 
$\mu_r$ dependence is greatly reduced at NLO as seen from the (hardly visible) 
width of the dark-shaded band in Fig.~\ref{fig:R_pm}.

We have compared our result for the Green function with a previous result 
from~\cite{Penin:1998ik,Penin:1998mx} and found agreement of the
scheme-independent terms at LO. The scheme-dependent finite-width  
terms have been fixed in these papers by matching the non-relativistic, 
resonant computation to the full theory diagram with an off-shell 
top-quark self-energy. This procedure accounts for part of the 
non-resonant contributions, but does not eliminate the need for a 
complete calculation of the $\gamma\gamma \to W^+ W^- b\bar b$ 
process with opposite photon helicities 
in the vicinity of $\sqrt{s}\approx 2 m_t$ to achieve 
parametric LO accuracy.\footnote{The corresponding calculations 
for $e^+ e^-\to t\bar {t}$ ~\cite{Beneke:2010mp,Penin:2011gg} suggest, 
however, that the contributions 
from the off-shell self-energy diagrams might be numerically the most 
important.}
Refs.~\cite{Penin:1998ik,Penin:1998mx} also 
present a calculation of the NLO P-wave Green function, but contrary 
to the closed expression (\ref{eq:delta1GP}) in dimensional regularization, 
the result is given in a sum representation that makes the comparison 
of even the scheme-independent terms difficult. We find, however, 
that we are able to 
reproduce the plots in~\cite{Penin:1998ik,Penin:1998mx} to a good
approximation, if we choose the scale of the hard matching coefficient
and $\mu_w$ at the hard scale.

\section{Stop-pair production at $e^+e^-$ and hadron
  colliders\label{sec:susy}}

The production of pairs of the scalar supersymmetric partner of the 
top quark is of interest at both hadron and $e^+ e^-$ colliders.
The threshold cross section of stop-antistop pairs in $e^+e^-$ collisions 
can be described in a fashion analogous to the previous 
sections~\cite{Bigi:1991mi}. Since 
the coupling of squarks to photons and $Z$ bosons contains a derivative, 
stops are produced in a P wave. We focus here solely on production of 
the lighter mass eigenstate $\tilde{t}\equiv\tilde{t}_1=
\tilde{t}_L\cos\theta_{\tilde{t}}+\tilde{t}_R\sin\theta_{\tilde{t}}$. The
ratio
$R_{\tilde{t}\tilde{\bar{t}}}=\sigma_{\tilde{t}\tilde{\bar{t}}X}/\sigma_0$
is given by 
\begin{equation}
 R_{\tilde{t}\tilde{\bar{t}}}=12\pi\,\Bigg(
e_{\tilde{t}}^2-\frac{2q^2}{q^2-M_Z^2}v_ez_{\tilde{t}}
e_{\tilde{t}}+\left(\frac{q^2}{q^2-M_Z^2}\right)^{\!2}\!
(v_e^2+a_e^2)z_{\tilde{t}}^2\Bigg)\text{ Im}\left[\Pi^{(\p)}(q^2)\right],
 \label{eq:Rratiostop}
\end{equation}
where $z_{\tilde{t}}$ is the coupling constant for the
$Z\tilde{t}\tilde{\bar{t}}$ vertex, which depends on the mixing 
angle $\theta_{\tilde{t}}$, and $\Pi^{(\p)}(q^2)$ is the
two-point function of the derivative current, which 
matches to the NRQCD current 
\begin{equation}
 j^{(\p)k}=\frac{1}{m_{\tilde t}}\, \psi^*\,i{\p}^k\chi^*.
\end{equation}
Here $\psi$ denotes the stop and $\chi$ the anti-stop field in the
non-relativistic normalization 
$\psi\sim\chi\sim m_{\tilde{t}}^{3/2}$. Including 
the hard matching coefficient, the two-point function takes the form
\begin{equation}
 \begin{aligned}
  \Pi^{(\p)}(q^2)=\frac{N_c}{4(d-1)m_{\tilde t}^4}\, c_{\p}^2\, G^P(E),
 \end{aligned}
 \label{eq:PipPNRQCD}
\end{equation}
where
\begin{equation}
 c_{\p}=1-4C_F\,\frac{\alpha_s}{4\pi}+\mathcal{O}(\alpha_s^2)
\end{equation}
is the hard matching coefficient of the current
$j^{(\p)k}$~\cite{Hoang:2005dk} and
\begin{equation}
 G^P(E)=\frac{i}{N_c}\int d^dx\,e^{iEx^0}
\Braket{0|T\left([\chi i{\p}^k\psi](x)[\psi^* i{\p}^k\chi^*](0)
\right)|0}_{\text{PNRQCD}}
\end{equation}
the P-wave Green function.
It is straightforward to evaluate this expression.
Since (\ref{eq:PipPNRQCD}) differs from the corresponding 
expression for top-antitop production only by the prefactor, we observe 
the same qualitative features. Most importantly, the theoretical 
uncertainty of the QCD contributions as measured by the residual 
$\mu_r$ dependence is greatly reduced at NLO as compared to the LO 
calculation~\cite{Bigi:1991mi}.

In the following we comment on the relevance of the P-wave for the 
production of stop-antistop pairs at hadron colliders. In quark-antiquark
annihilation the t-channel  gluino exchange
diagram that is dominant for pair production of
light-flavour squarks is strongly suppressed for stops due to the
negligible top parton distribution function in the proton. 
Thus, in quark-antiquark annihilation the s-channel, 
which produces stop pairs in a P-wave and
colour-octet state, is the dominant contribution. Production of stop-antistop 
pairs in gluon-fusion can be described as for light-flavour
squarks. In~\cite{Beneke:2010da} a formalism for resummation of
soft and Coulomb corrections in S-wave production of pairs of heavy
coloured particles was derived at the next-to-next-to-leading logarithmic 
approximation (NNLL). This was generalized to stop
pair production at NLL in~\cite{Falgari:2012hx,Falgari:2012sq}. 
The arguments presented
there suggest that the factorization formula also holds at NNLL. The
NLO P-wave Green function derived in this work constitutes the
dominant part of the potential function $J^{R_\alpha}$ accounting for
NNLL terms beginning with $\alpha_s^2/\beta$. To obtain the colour-octet
Green function one only has to make the replacement $-C_F\ra
D_8=1/(2N_c)$ in (\ref{eq:VCoulombcoefficients}). Additionally, 
non-Coulomb potentials yield
terms beginning with $\alpha_s^2\log\beta$~\cite{Beneke:2013opa}, 
which also have to be included at NNLL order.

\section{Conclusion\label{sec:conclusion}}

We have computed the P-wave Green function in dimensional
regularization up to NLO. We further confirmed results for the NLO
correction to energy levels and wave functions at the origin of 
P-wave bound states. The NLO correction reduces the renormalization 
scale uncertainty considerably. We have discussed the P-wave 
contribution to three different pair production
processes. The NLO P-wave contribution to the top-quark pair production cross
section near threshold is part of the complete NNNLO result. 
The P-wave production cross section turns out to be small relative 
to the dominant S-wave, below 1\%. It is included in the forth-coming 
NNNLO result for the $e^+e^-\ra t\bar{t}X$ cross section~\cite{Beneke:2014}. 
The photon collider option further offers the possibility to produce 
tops in a pure P-wave and with a larger cross section.

In $e^+e^-$ collisions squark-antisquark pairs are also produced in a
P-wave. We have given the necessary formulas for the NLO cross
section. If squarks that are sufficiently light for production at a
future linear collider should be found, this will allow precision
studies including a precise mass determination. The NLO P-wave Green
function is also an important ingredient in the NNLL prediction of
stop-antistop production in hadron collisions.

\noindent
\subsubsection*{Acknowledgement}
This work has been supported by the DFG
Sonder\-forschungs\-bereich/Transregio~9 ``Com\-pu\-ter\-gest\"utzte
Theoretische Teil\-chen\-physik'', the Gottfried Wilhelm Leibniz 
programme of the Deutsche Forschungsgemeinschaft (DFG), 
and the DFG cluster of excellence ``Origin and Structure of the Universe.''

\appendix

\section{Computation of the diagrams \label{sec:diagrams}}

We compute the divergent part of the single-insertion function~\eqref{eq:IPa}
in momentum space. The integrals contain only a single scale $mE$, which
appears as a mass term in the non-relativistic heavy quark propagators 
after performing the integrations over the zero-component of the loop momenta.
We rescale the integration momenta by this scale,  
$\mathbf{p}\ra\sqrt{-mE}\,\mathbf{k}$, to make them dimensionless.
We then use FIRE~\cite{Smirnov:2008iw,Smirnov:2013dia} to reduce the diagrams
to a set of master integrals. In the master integrals solid lines denote
rescaled massive propagators, which take the form $1/(\mathbf{k}^2+1)$.
Dashed lines denote potential gluons and wavy lines insertions of the NLO
Coulomb potential, i.e. gluon propagators with an index $1+u$.
Initially, we make the assignment $\mu_r$ to all $\mu$ raised to powers
of $u$ and $\mu_w$ to all $\mu$ raised to powers of $\epsilon$, since the
former arise from the running coupling contribution to the NLO Coulomb
potential. There are however some subtleties associated with this scale
separation, which will be discussed below. We obtain
\begin{equation}
 I_{\text{P}}^{00}[1+u]=m^3E\left(\frac{-mE}{\mu_r^2}\right)^{-u}\left(\frac{-mE}{\tilde{\mu}_w^2}\right)^{-2\epsilon}
 \left[\frac{1+u}{1-u-2\epsilon}\,\vcenter{\hbox{\includegraphics[height=1cm]{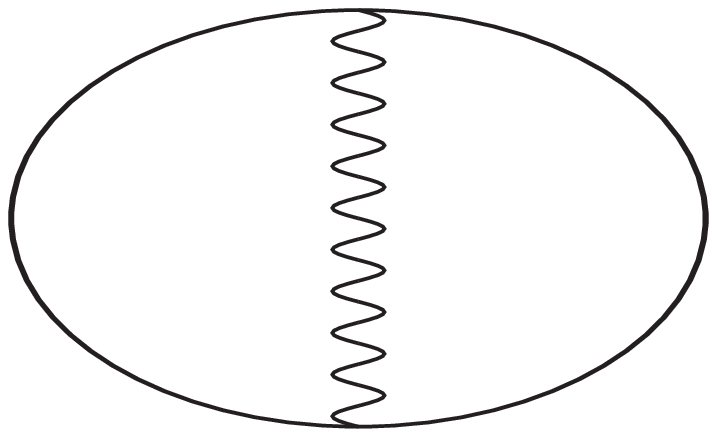}}}\right],
\label{eq:firstdiagram}
\end{equation}
\begin{eqnarray}
&&I_{\text{P}}^{10}[1+u]=m^3E\lambda\left(\frac{-mE}{\mu_r^2}\right)^{-u}\left(\frac{-mE}{\tilde{\mu}_w^2}\right)^{-3\epsilon}\,(8\pi)
\nonumber\\
&&\hspace*{0.0cm} 
\times\,\Bigg[\frac{1+u}{4 \epsilon  (1-u-2\epsilon)}\,\vcenter{\hbox{\includegraphics[height=1cm]{fig_F111.eps}}}\vcenter{\hbox{\includegraphics[height=1cm]{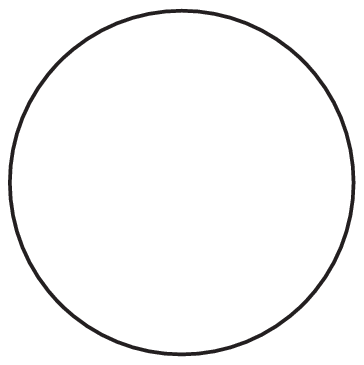}}}
 -\frac{1-u-4 \epsilon-2 u^2-6u\epsilon-4\epsilon^2}{8 \epsilon  (1-u-2\epsilon)}\,\vcenter{\hbox{\includegraphics[height=1cm]{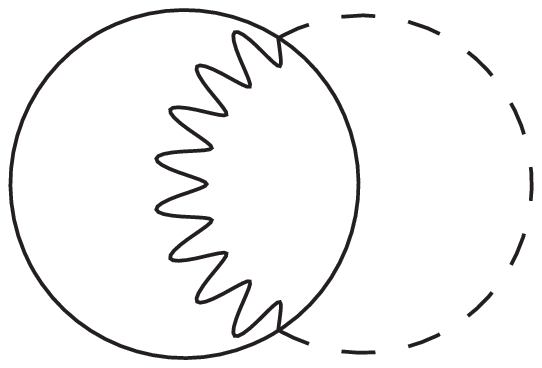}}}\Bigg],
\qquad
\end{eqnarray}
\begin{eqnarray}
 &&I_{\text{P}}^{20}[1+u]=-m^3E\lambda^2\left(\frac{-mE}{\mu_r^2}\right)^{-u}\left(\frac{-mE}{\tilde{\mu}_w^2}\right)^{-4\epsilon}\frac{(8\pi)^2}{64 \epsilon ^2 (2 \epsilon -1) (u+2 \epsilon -1) (u+3 \epsilon )}\nonumber \\
 &&\times\Bigg\{(u+2 \epsilon ) (u+4 \epsilon -1) \Big[u^2 (4 \epsilon -2)+u (4 \epsilon  (7 \epsilon -1)-1)
\nonumber \\
&&\hspace{4.4cm} +\,4 \epsilon  \left(12 \epsilon ^2+\epsilon -1\right)+1\Big]
 \,\vcenter{\hbox{\includegraphics[height=1cm]{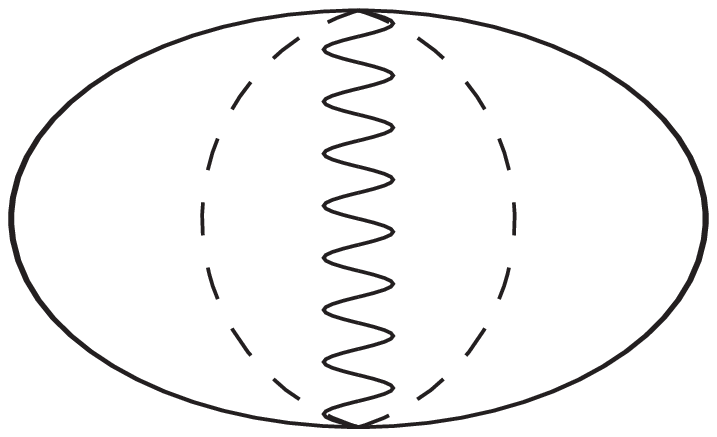}}}
\nonumber \\
 &&\hspace{0.5cm}+\,2(u+3 \epsilon)\Bigg[(2 \epsilon -1) \left(2 u^2+6 u \epsilon +u+4 \epsilon  (\epsilon +1)-1\right)\,\vcenter{\hbox{\includegraphics[height=1cm]{fig_F10111.eps}}}
 \vcenter{\hbox{\includegraphics[height=1cm]{fig_F1.eps}}}\nonumber \\
 &&\hspace{0.5cm}+\,2 (1 + u)\Big[-2 \epsilon  (4 \epsilon  (\epsilon +1)-1)\,\vcenter{\hbox{\includegraphics[height=1cm]{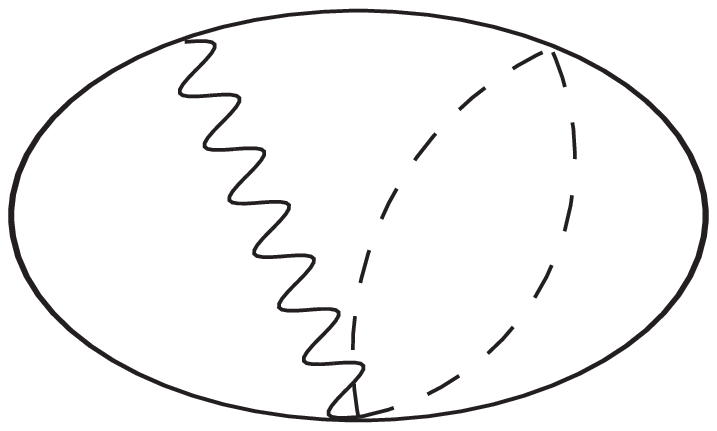}}}+(2 \epsilon -1)
 \,\vcenter{\hbox{\includegraphics[height=1cm]{fig_F111.eps}}}\left(\vcenter{\hbox{\includegraphics[height=1cm]{fig_F1.eps}}}\right)^2\Big]\Bigg]\Bigg\},\;\;
\qquad
\label{eq:thirddiagram}
\end{eqnarray}
\begin{eqnarray}
 &&I_{\text{P}}^{11}[1+u]=-m^3E\lambda^2\left(\frac{-mE}{\mu_r^2}\right)^{-u}\left(\frac{-mE}{\tilde{\mu}_w^2}\right)^{-4\epsilon}
 \frac{(8\pi)^2}{32 \epsilon ^2 (u+2 \epsilon -1) (u+3 \epsilon )}\nonumber\\
 &&\times\,\Bigg\{(u+4 \epsilon -1) \left[u \left(2 u^2+u-1\right)+8 (4 u+1) \epsilon ^2+2 (u (7 u+3)-2) \epsilon +24 \epsilon ^3\right]\,\vcenter{\hbox{\includegraphics[height=1cm]{fig_F1001111.eps}}}\nonumber\\
 &&\hspace{0.5cm}+\,2(u+3 \epsilon)\Bigg[\left(2 u^2+6 u \epsilon +u+4 \epsilon  (\epsilon +1)-1\right)\,\vcenter{\hbox{\includegraphics[height=1cm]{fig_F10111.eps}}}
 \vcenter{\hbox{\includegraphics[height=1cm]{fig_F1.eps}}}\nonumber\\
 &&\hspace{0.5cm}+\,(u+1)\,\vcenter{\hbox{\includegraphics[height=1cm]{fig_F111.eps}}}\left(\vcenter{\hbox{\includegraphics[height=1cm]{fig_F1.eps}}}\right)^2\Bigg]\Bigg\}.
\label{eq:fourthdiagram}
\end{eqnarray}
The master diagrams with just two massive lines can be computed with standard
methods. We find
\begin{equation}
 \vcenter{\hbox{\includegraphics[height=1cm]{fig_F1.eps}}}\equiv\int\frac{d^{d-1}\mathbf{k}}{(2\pi)^{d-1}}\frac{1}{\mathbf{k}^2+1}=\frac{1}{(4\pi)^{\frac{3}{2}-\epsilon}}\Gamma\left(-\frac{1}{2}+\epsilon\right),
\end{equation}
\begin{equation}
\begin{aligned}
 \vcenter{\hbox{\includegraphics[height=1cm]{fig_F111.eps}}}\equiv&\int\frac{d^{d-1}\mathbf{k}_1}{(2\pi)^{d-1}}\int\frac{d^{d-1}\mathbf{k}_1}{(2\pi)^{d-1}}
 \frac{1}{[\mathbf{k}_1^2+1][\mathbf{k}_2^2+1][(\mathbf{k}_1-\mathbf{k}_2)^2]^{1+u}}\\
 =&\frac{1}{(4\pi)^{3-2\epsilon}}(-2\pi)
 \frac{\Gamma(-2u-2\epsilon)\Gamma(1/2+u+\epsilon)\sin[\pi(u+\epsilon)]}{\Gamma(1-u-2\epsilon)\Gamma(3/2-\epsilon)\sin[\pi(u+2\epsilon)]},
 \end{aligned}
\end{equation}
\begin{eqnarray}
 \vcenter{\hbox{\includegraphics[height=1cm]{fig_F10111.eps}}}&\equiv
&\int\left[\,\prod\limits_{j=1}^3\frac{d^{d-1}\mathbf{k}_j}{(2\pi)^{d-1}}\right]
 \frac{1}{[\mathbf{k}_1^2+1][\mathbf{k}_3^2+1][(\mathbf{k}_1-\mathbf{k}_2)^2]^{1+u}(\mathbf{k}_2-\mathbf{k}_3)^2}\nonumber \\
 &=&\frac{1}{(4\pi)^{\frac{9}{2}-3\epsilon}}
 \frac{-2^{1-2 u-4 \epsilon} \pi ^{3/2}\Gamma \left(1/2-u-2 \epsilon\right) \Gamma (u+2 \epsilon )}{(1-2 \epsilon)\Gamma(1+u)\Gamma\left(3/2-u-3\epsilon\right)\cos(\pi\epsilon)}\nonumber\\
 &&\times\,\left(\frac{1}{\cos[\pi(u+\epsilon)]}+\frac{1}{\cos[\pi(u+3\epsilon)]}\right),
\end{eqnarray}
\begin{equation}
\begin{aligned}
 \vcenter{\hbox{\includegraphics[height=1cm]{fig_F1001111.eps}}}\equiv&
\int\left[\,\prod\limits_{j=1}^4\frac{d^{d-1}\mathbf{k}_j}{(2\pi)^{d-1}}\right]
 \frac{1}{[\mathbf{k}_1^2+1][\mathbf{k}_4^2+1][(\mathbf{k}_1-\mathbf{k}_2)^2]^{1+u}(\mathbf{k}_2-\mathbf{k}_3)^2(\mathbf{k}_3-\mathbf{k}_4)^2}\\
 =&\frac{1}{(4\pi)^{6-4\epsilon}}\left(\frac{1}{\text{sin}[\pi(u+2\epsilon)]}+\frac{1}{\text{sin}[\pi(u+4\epsilon)]}\right)\left(
  -2^{2-2u-6\epsilon}\pi^{3/2} \right)\\
 &\times\frac{\Gamma\left(1/2-\epsilon\right)\Gamma(1-u-3\epsilon)\Gamma\left(1/2-u-\epsilon\right)\Gamma\left(-1/2+u+3\epsilon\right)}
 {(1-2\epsilon)\Gamma(1+u)\Gamma(2-u-4\epsilon)\Gamma(1-u-2\epsilon)\cos(\pi\epsilon)}.
 \end{aligned}
\end{equation}
The remaining master integral contains three massive lines and is
therefore more complicated:
\begin{eqnarray}
  \lefteqn{\vcenter{\hbox{\includegraphics[height=1cm]{fig_F1101111.eps}}}}
  && \nonumber\\
  &\equiv& \int\left[\,\prod\limits_{j=1}^4\frac{d^{d-1}\mathbf{k}_j}{(2\pi)^{d-1}}
\right]\!
 \frac{1}{[\mathbf{k}_1^2+1][\mathbf{k}_2^2+1][\mathbf{k}_4^2+1][(\mathbf{k}_1-\mathbf{k}_2)^2]^{1+u}(\mathbf{k}_2-\mathbf{k}_3)^2(\mathbf{k}_3-\mathbf{k}_4)^2}. \nonumber\\
\end{eqnarray}
The solution for $u=0$ can be found in~\cite{Schroder:2005va} and
agrees with our result obtained by using FIRE and the known master
integrals. For $u=\epsilon$ we calculate the diagram as an expansion
in $\epsilon$ up to order $\epsilon^2$. The quadratic term is
required, because the coefficient of this integral
in~\eqref{eq:thirddiagram} and the potential each contain a factor
$\epsilon^{-1}$. We use the MB.m package~\cite{Czakon:2005rk} to
perform an analytic continuation of the Mellin-Barnes integral in
\(\epsilon\). We then close the integration contours to pick up single
and double infinite sums over the residues of the integrand. These
sums can be transformed into cyclotomic harmonic sums. They were
reduced to a set of known basis sums using the \textit{Harmonic Sums}
package~\cite{Ablinger:2010kw,Ablinger:2013hcp,Ablinger:2013cf,Ablinger:2011te,Blumlein:2009ta,Remiddi:1999ew,Vermaseren:1998uu}. The
result is given by 
\begin{eqnarray}
  \vcenter{\hbox{\includegraphics[height=1cm]{fig_F1101111.eps}}}\Big|_{u=\epsilon} & = & \frac{\exp(-4\epsilon\gamma_E)}{(4\pi)^{6-4\epsilon}}
  \Bigg[\frac{2 \pi ^2}{5 \epsilon ^2}+\frac{4 \pi ^2}{5\epsilon}(4-5 \ln2)+\frac{4\pi ^2}{5}\left(26-40\ln2+25\ln^2 2\right)\nonumber\\
  & & +\frac{8\pi ^2}{15}\epsilon\left(240-390 \ln2+300\ln^2 2-125\ln^3 2+32 \zeta (3)\right)\nonumber\\
  & & +\frac{4\pi^2}{225}\epsilon ^2\Big(43560-326 \pi ^4-72000\ln2+58500 \ln^2 2\nonumber\\
  & & -30000\ln^3 2+9375\ln^4 2+1920(4-5\ln2)\zeta(3)\Big)+\mathcal{O}(\epsilon^3)\Bigg].
\end{eqnarray}
Given these master integrals we obtain the results for the diagrams
by means of~\eqref{eq:firstdiagram}--\eqref{eq:fourthdiagram}.
Note, however, that the choice $u=\epsilon$ in~\eqref{eq:firstdiagram}
--\eqref{eq:fourthdiagram} results in spurious logarithms
$\ln(\mu_r/\mu_w)$ that arise from terms of the form
\begin{equation}
 \frac{1}{n\epsilon+u}\mu_r^u\mu_w^{n\epsilon}\,\mathop{=}\limits^{u=\epsilon}
\,\frac{1}{(n+1)\epsilon}+\ln(\mu_w)+\frac{1}{n+1}\ln\left(\frac{\mu_r}{\mu_w}\right)
+\mathcal{O}(\epsilon).
\end{equation}
This is due to the fact that the origin of poles cannot be unambiguously 
identified in dimensionally regulated multi-loop integrals. To obtain the 
correct scale assignment, we subtract from
the results of~\eqref{eq:firstdiagram}--\eqref{eq:fourthdiagram} the
scale dependent logarithms and then add the respective terms obtained
without the identification $u=\epsilon$, i.e. with the Coulomb potential
expanded in $\epsilon$. Furthermore, the scales $\mu_r$ and $\mu_w$ are
set equal in pole terms  $\ln(\mu_r/\mu_w)/\epsilon$ to ensure that the 
pole terms are scale-independent.

The results for the individual diagrams are:
\begin{eqnarray}
  I_{\text{P}}^{00}[1] & = &
  \frac{m^3E}{(4\pi)^2}\left[\frac{1}{4 \epsilon}
    +1+L_\lambda^w\right], \nonumber \\
  I_{\text{P}}^{00}[1+\epsilon] & = &
  \frac{m^3E}{(4\pi)^2}\left[\frac{1}{6 \epsilon }+1+L_\lambda^w
  \right], \nonumber \\
  \frac{1}{\epsilon}\left[I_{\text{P}}^{00}[1+\epsilon]-I_{\text{P}}^{00}[1]\right]
  & = & 
  \frac{m^3E}{(4\pi)^2}\left[-\frac{1}{12 \epsilon ^2} +
    1+\frac{17 \pi ^2}{72}
    +2L_\lambda^r+2L_\lambda^rL_\lambda^w
    -\left(L_\lambda^w\right)^2\right].\qquad
 \label{eq:I00}
\\
I_{\text{P}}^{10}[1]=I_{\text{P}}^{10}[1+\epsilon]&=&\frac{m^3E}{(4\pi)^2}\lambda\left[\frac12+\frac{\pi^2}{6}\right],\nonumber\\
\frac{1}{\epsilon}\left[I_{\text{P}}^{10}[1+\epsilon]-I_{\text{P}}^{10}[1]\right]&=&\frac{m^3E}{(4\pi)^2}\lambda
\left[\frac{5}{2}+\frac{\pi ^2}{3}-2 \zeta (3)+\left(1+\frac{\pi ^2}{3}\right)L_\lambda^r\right].
  \label{eq:I10}
\\
I_{\text{P}}^{20}[1]=I_{\text{P}}^{11}[1]&=&\frac{m^3E}{(4\pi)^2}\lambda^2\left[-\frac{1}{8\epsilon}-\frac{7}{4}+\zeta(3)-L_\lambda^w\right],\nonumber\\
I_{\text{P}}^{20}[1+\epsilon]&=&\frac{m^3E}{(4\pi)^2}\lambda^2\left[-\frac{1}{10
    \epsilon }-\frac{17}{10}+\zeta (3)-L_\lambda^w \right],\nonumber\\
\frac{1}{\epsilon}\left[I_{\text{P}}^{20}[1+\epsilon]-I_{\text{P}}^{20}[1]\right]&=&
\frac{m^3E}{(4\pi)^2}\lambda^2\bigg[\frac{1}{40 \epsilon^2}
+\frac{1}{20 \epsilon }
-\frac{29}{10}-\frac{\pi ^2}{24}-\frac{\pi ^4}{180}
+2 \zeta(3)+\frac12L_\lambda^w\nonumber\\
&&\phantom{\frac{m^3E}{(4\pi)^2}\lambda^2\bigg[}
-\left(\frac72-2\zeta(3)\right)L_\lambda^r+\left(L_\lambda^w\right)^2
-2L_\lambda^wL_\lambda^r \bigg].
  \label{eq:I20}
\\
I_{\text{P}}^{11}[1+\epsilon]&=&
\frac{m^3E}{(4\pi)^2}\lambda^2\left[-\frac{1}{10 \epsilon }
-\frac{9}{5}+\zeta (3)-L_\lambda^w \right],\nonumber\\
\frac{1}{\epsilon}\left[I_{\text{P}}^{11}[1+\epsilon]-I_{\text{P}}^{11}[1]\right]&=&
\frac{m^3E}{(4\pi)^2}\lambda^2\bigg[\frac{1}{40 \epsilon^2}-\frac{1}{20 \epsilon }
-\frac{29}{5}+\frac{\pi ^2}{8}+\frac{\pi ^4}{180}+2 \zeta(3)-\frac12L_\lambda^w\nonumber\\
&&\phantom{\frac{m^3E}{(4\pi)^2}\lambda^2\bigg[}-\left(\frac72-2\zeta(3)\right)L_\lambda^r+\left(L_\lambda^w\right)^2
-2L_\lambda^wL_\lambda^r \bigg].
  \label{eq:I11}
\end{eqnarray}

\section{Evaluation of the hypergeometric function \label{sec:pFq}}

The generalized hypergeometric function in the result for NLO Green
function can be expressed in terms of harmonic sums. This is useful
when our result is applied to particles with vanishing width, which
means that $\lambda$ has a large positive real part as one approaches
the threshold from below.
The necessary analytic continuation can be easily done for
the harmonic sums, see for example~\cite{Blumlein:2009ta,Albino:2009ci}.
We first use
\begin{eqnarray}
  _4F_3(1,1,4,4;5,5,3-\lambda;1)
  &=& 4 (\lambda -2)(\lambda-1)\lambda(\lambda+1)\Bigg[\frac{2}{3(1+\lambda)}\,
  _4F_3(1,1,1,1;2,2,-\lambda;1) \nonumber \\
  && \hspace*{-3cm} 
+\,\frac{1}{27}\left(\frac{\lambda  (\lambda  (3 (8-17 \lambda )
      \lambda -20)+11)-18}{(\lambda -1)^2 \lambda ^2 (\lambda
      +1)}-33\psi_1(2-\lambda)\right)\Bigg] 
\end{eqnarray}
to change the arguments of the function to more suitable values.
Following Appendix A.1 of~\cite{Beneke:2011mq}, the remaining
hypergeometric function can be rewritten as the Mellin transform of a
dilogarithm, which can further be expressed through harmonic sums
\begin{eqnarray}
 &&\frac{1}{(1+\lambda)}\, _4F_3(1,1,1,1;2,2,-\lambda;1)=-\mathbf{M}\left[\frac{\text{Li}_2(1-x)}{1-x}\right](-2-\lambda) \nonumber \\
&& =\,-\left[S_1(-2-\lambda)S_2(-2-\lambda)-\zeta(2) S_1(-2-\lambda)+S_3(-2-\lambda)-S_{2,1}(-2-\lambda)+\zeta(3)\right].
\nonumber\\[-0.1cm]
&&\end{eqnarray}
The latter step was performed with the help of the
\texttt{FORM}~\cite{Vermaseren:2000nd} program
\texttt{HARMPOL}~\cite{Remiddi:1999ew}. The (nested) harmonic sums are
defined as $S_a(N) = \sum_{i=1}^N\frac{1}{i^a}$ and $S_{a,b}(N) =
\sum_{i=1}^N\frac{1}{i^a}\,S_b(i)$.

\bibliographystyle{JHEP-2}

\begin{thebibliography}{10%
0}

\bibitem{Martinez:2002st}
  M.~Martinez and R.~Miquel,
  Eur.\ Phys.\ J.\ C {\bf 27} (2003) 49
  [hep-ph/0207315].

\bibitem{Seidel:2013sqa}
  K.~Seidel, F.~Simon, M.~Tesa\v{r} and S.~Poss,
  Eur.\ Phys.\ J.\ C {\bf 73} (2013) 2530,
  arXiv:1303.3758 [hep-ex].

\bibitem{Horiguchi:2013wra}
  T.~Horiguchi, A.~Ishikawa, T.~Suehara, K.~Fujii, Y.~Sumino, Y.~Kiyo and H.~Yamamoto,
  arXiv:1310.0563 [hep-ex].

\bibitem{Thacker:1990bm}
  B.~A.~Thacker and G.~P.~Lepage,
  Phys.\ Rev.\ D {\bf 43} (1991) 196.

\bibitem{Lepage:1992tx}
  G.~P.~Lepage, L.~Magnea, C.~Nakhleh, U.~Magnea and K.~Hornbostel,
  Phys.\ Rev.\ D {\bf 46} (1992) 4052
  [hep-lat/9205007].

\bibitem{Bodwin:1994jh}
  G.~T.~Bodwin, E.~Braaten and G.~P.~Lepage,
  Phys.\ Rev.\ D {\bf 51} (1995) 1125
   [Erratum-ibid.\ D {\bf 55} (1997) 5853]
  [hep-ph/9407339].

\bibitem{Pineda:1997bj}
  A.~Pineda and J.~Soto,
  Nucl.\ Phys.\ Proc.\ Suppl.\  {\bf 64} (1998) 428
  [hep-ph/9707481].

\bibitem{Pineda:1997ie}
  A.~Pineda and J.~Soto,
  Phys.\ Lett.\ B {\bf 420} (1998) 391
  [hep-ph/9711292].

\bibitem{Beneke:1998jj}
  M.~Beneke,
  in: Proceedings of the 33rd Rencontres de Moriond: Electroweak 
  Interactions and Unified Theories, Les Arcs, France, 14-21 Mar 1998, 
  hep-ph/9806429.

\bibitem{Beneke:1999qg}
M.~Beneke, A.~Signer and V.~A. Smirnov, 
Phys.\ Lett.\ B {\bf 454} (1999) 137 [hep-ph/9903260].

\bibitem{Brambilla:1999xf}
  N.~Brambilla, A.~Pineda, J.~Soto and A.~Vairo,
  Nucl.\ Phys.\ B {\bf 566} (2000) 275
  [hep-ph/9907240].

\bibitem{Beneke:2013PartI}
  M.~Beneke, Y.~Kiyo and K.~Schuller,
  {\em ``Third-order correction to top-quark pair production near
  threshold I. Effective theory set-up and matching coefficients,''}
  preceding article.

\bibitem{Beneke:2005hg}
  M.~Beneke, Y.~Kiyo and K.~Schuller,
  Nucl.\ Phys.\ B {\bf 714} (2005) 67
  [hep-ph/0501289].



\bibitem{Beneke:2008ec}
  M.~Beneke, Y.~Kiyo and K.~Schuller,
  PoS RADCOR {\bf 2007} (2007) 051,
  arXiv:0801.3464 [hep-ph].

\bibitem{Beneke:2008cr}
  M.~Beneke and Y.~Kiyo,
  Phys.\ Lett.\ B {\bf 668} (2008) 143,
  arXiv:0804.4004 [hep-ph].

\bibitem{Beneke:2013PartII}
  M.~Beneke, Y.~Kiyo and K.~Schuller,
  {\em ``Third-order correction to top-quark pair production near
  threshold II. Potential contributions,''}
in preparation.

\bibitem{Bigi:1991mi}
  I.~I.~Y.~Bigi, V.~S.~Fadin and V.~A.~Khoze,
  Nucl.\ Phys.\ B {\bf 377} (1992) 461.
  
\bibitem{Penin:1998ik}
  A.~A.~Penin and A.~A.~Pivovarov,
  Nucl.\ Phys.\ B {\bf 550} (1999) 375
  [hep-ph/9810496].
  
\bibitem{Penin:1998mx}
  A.~A.~Penin and A.~A.~Pivovarov,
  Phys.\ Atom.\ Nucl.\  {\bf 64} (2001) 275
   [Yad.\ Fiz.\  {\bf 64} (2001) 323]
  [hep-ph/9904278].
  
\bibitem{Kuhn:1999hw}
  J.~H.~K{\"u}hn and T.~Teubner,
  Eur.\ Phys.\ J.\ C {\bf 9} (1999) 221
  [hep-ph/9903322].

\bibitem{Beneke:2003xh}
  M.~Beneke, A.~P.~Chapovsky, A.~Signer and G.~Zanderighi,
  Phys.\ Rev.\ Lett.\  {\bf 93} (2004) 011602
  [hep-ph/0312331].

\bibitem{Beneke:2004km}
  M.~Beneke, A.~P.~Chapovsky, A.~Signer and G.~Zanderighi,
  Nucl.\ Phys.\ B {\bf 686} (2004) 205
  [hep-ph/0401002].

\bibitem{Beneke:2010mp}
  M.~Beneke, B.~Jantzen and P.~Ruiz-Femen\'{i}a,
  Nucl.\ Phys.\ B {\bf 840} (2010) 186,
  arXiv:1004.2188 [hep-ph].

\bibitem{Jantzen:2013gpa}
  B.~Jantzen and P.~Ruiz-Femen\'{i}a,
  Phys.\ Rev.\ D {\bf 88} (2013) 054011,
  arXiv:1307.4337 [hep-ph].

\bibitem{Kniehl:2006qw}
  B.~A.~Kniehl, A.~Onishchenko, J.~H.~Piclum and M.~Steinhauser,
  Phys.\ Lett.\ B {\bf 638} (2006) 209
  [hep-ph/0604072].

\bibitem{Beneke:2010da}
  M.~Beneke, P.~Falgari and C.~Schwinn,
  Nucl.\ Phys.\ B {\bf 842} (2011) 414,
  arXiv:1007.5414 [hep-ph].

\bibitem{Wichmann:1961}
  E.~Wichmann and C.~Woo,
  J.\ Math.\ Ph. {\bf 2} (1961) 178.

\bibitem{Schwinger:1964zzb}
  J.~Schwinger,
  J.\ Math.\ Phys.\  {\bf 5} (1964) 1606.
  
\bibitem{Voloshin:1985bd}
  M.~B.~Voloshin,
  Sov.\ J.\ Nucl.\ Phys.\  {\bf 40} (1984) 662
   [Yad.\ Fiz.\  {\bf 40} (1984) 1039].
  
\bibitem{Voloshin:1979uv}
  M.~B.~Voloshin,
  Sov.\ J.\ Nucl.\ Phys.\  {\bf 36} (1982) 143
   [Yad.\ Fiz.\  {\bf 36} (1982) 247].
  
\bibitem{Fadin:1987wz}
  V.~S.~Fadin and V.~A.~Khoze,
  JETP Lett.\  {\bf 46} (1987) 525
   [Pisma Zh.\ Eksp.\ Teor.\ Fiz.\  {\bf 46} (1987) 417].
  
\bibitem{Fadin:1988fn}
  V.~S.~Fadin and V.~A.~Khoze,
  Sov.\ J.\ Nucl.\ Phys.\  {\bf 48} (1988) 309
   [Yad.\ Fiz.\  {\bf 48} (1988) 487].
  
\bibitem{Hoang:2004tg}
  A.~H.~Hoang and C.~J.~Rei\ss er,
  Phys.\ Rev.\ D {\bf 71} (2005) 074022
  [hep-ph/0412258].
  
\bibitem{Hoang:2010gu}
  A.~H.~Hoang, C.~J.~Rei\ss er and P.~Ruiz-Femen\'{i}a,
  Phys.\ Rev.\ D {\bf 82} (2010) 014005,
  arXiv:1002.3223 [hep-ph].

\bibitem{Smirnov:2008iw}
  A.~V.~Smirnov,
  JHEP {\bf 0810} (2008) 107,
  arXiv:0807.3243 [hep-ph].

\bibitem{Smirnov:2013dia}
  A.~V.~Smirnov and V.~A.~Smirnov,
  arXiv:1302.5885 [hep-ph].
  
\bibitem{Schroder:2005va}
  Y.~Schr{\"o}der and A.~Vuorinen,
  JHEP {\bf 0506} (2005) 051
  [hep-ph/0503209].

\bibitem{Hoang:2001mm}
  A.~H.~Hoang, A.~V.~Manohar, I.~W.~Stewart and T.~Teubner,
  Phys.\ Rev.\ D {\bf 65} (2002) 014014
  [hep-ph/0107144].

\bibitem{Beneke:1994sw}
  M.~Beneke and V.~M.~Braun,
  Nucl.\ Phys.\ B {\bf 426} (1994) 301
  [hep-ph/9402364].
  
\bibitem{Bigi:1994em}
  I.~I.~Y.~Bigi, M.~A.~Shifman, N.~G.~Uraltsev and A.~I.~Vainshtein,
  Phys.\ Rev.\ D {\bf 50} (1994) 2234
  [hep-ph/9402360].
  
\bibitem{Beneke:1994rs}
  M.~Beneke,
  Phys.\ Lett.\ B {\bf 344} (1995) 341
  [hep-ph/9408380].
  
\bibitem{Beneke:1998rk}
  M.~Beneke,
  Phys.\ Lett.\ B {\bf 434} (1998) 115
  [hep-ph/9804241].
  
\bibitem{Ginzburg:1982yr}
  I.~F.~Ginzburg, G.~L.~Kotkin, S.~L.~Panfil, V.~G.~Serbo and V.~I.~Telnov,
  Nucl.\ Instrum.\ Meth.\ A {\bf 219} (1984) 5.

\bibitem{Badelek:2001xb}
  B.~Badelek {\it et al.}  [ECFA/DESY Photon Collider Working Group Collaboration],
  Int.\ J.\ Mod.\ Phys.\ A {\bf 19} (2004) 5097
  [hep-ex/0108012].

\bibitem{Adolphsen:2013kya}
  C.~Adolphsen, M.~Barone, B.~Barish, K.~Buesser, P.~Burrows, J.~Carwardine, J.~Clark and H\'{e}l\`{e}ne~M.~Durand {\it et al.},
  arXiv:1306.6328 [physics.acc-ph].

\bibitem{Penin:2011gg}
  A.~A.~Penin and J.~H.~Piclum,
  JHEP {\bf 1201} (2012) 034,
  arXiv:1110.1970 [hep-ph].

\bibitem{Hoang:2005dk}
  A.~H.~Hoang and P.~Ruiz-Femen\'{i}a,
  Phys.\ Rev.\ D {\bf 73} (2006) 014015
  [hep-ph/0511102].

\bibitem{Falgari:2012hx}
  P.~Falgari, C.~Schwinn and C.~Wever,
  JHEP {\bf 1206} (2012) 052,
  arXiv:1202.2260 [hep-ph].

\bibitem{Falgari:2012sq}
  P.~Falgari, C.~Schwinn and C.~Wever,
  JHEP {\bf 1301} (2013) 085,
  arXiv:1211.3408 [hep-ph].

\bibitem{Beneke:2013opa}
  M.~Beneke, P.~Falgari, J.~Piclum, C.~Schwinn and C.~Wever,
  arXiv:1312.0837 [hep-ph].

\bibitem{Beneke:2014}
  M.~Beneke, Y.~Kiyo, P.~Marquard, A.~Penin, J.~Piclum, K~Schuller,
  D.~Seidel and M.~Steinhauser,
  in preparation.

\bibitem{Czakon:2005rk}
  M.~Czakon,
  Comput.\ Phys.\ Commun.\  {\bf 175} (2006) 559
  [hep-ph/0511200].

\bibitem{Ablinger:2010kw}
  J.~Ablinger,
  arXiv:1011.1176 [math-ph].

\bibitem{Ablinger:2013hcp}
  J.~Ablinger,
  arXiv:1305.0687 [math-ph].

\bibitem{Ablinger:2013cf}
  J.~Ablinger, J.~Bl{\"u}mlein and C.~Schneider,
  J.\ Math.\ Phys.\  {\bf 54} (2013) 082301,
  arXiv:1302.0378 [math-ph].
  
\bibitem{Ablinger:2011te}
  J.~Ablinger, J.~Bl{\"u}mlein and C.~Schneider,
  J.\ Math.\ Phys. {\bf 52} (2011) 102301,
  arXiv:1105.6063 [math-ph].

\bibitem{Blumlein:2009ta}
  J.~Bl{\"u}mlein,
  Comput.\ Phys.\ Commun.\  {\bf 180} (2009) 2218,
  arXiv:0901.3106 [hep-ph].

\bibitem{Remiddi:1999ew}
  E.~Remiddi and J.~A.~M.~Vermaseren,
  Int.\ J.\ Mod.\ Phys.\ A {\bf 15} (2000) 725
  [hep-ph/9905237].

\bibitem{Vermaseren:1998uu}
  J.~A.~M.~Vermaseren,
  Int.\ J.\ Mod.\ Phys.\ A {\bf 14} (1999) 2037
  [hep-ph/9806280].

\bibitem{Albino:2009ci}
  S.~Albino,
  Phys.\ Lett.\ B {\bf 674} (2009) 41,
  arXiv:0902.2148 [hep-ph].

\bibitem{Beneke:2011mq}
  M.~Beneke, P.~Falgari, S.~Klein and C.~Schwinn,
  Nucl.\ Phys.\ B {\bf 855} (2012) 695,
  arXiv:1109.1536 [hep-ph].

\bibitem{Vermaseren:2000nd}
  J.~A.~M.~Vermaseren,
  math-ph/0010025.

\end{thebibliography}

\providecommand{\href}[2]{#2}\begingroup\raggedright\endgroup

\end{document}